\documentstyle[prl,aps,preprint]{revtex}

\begin{document}

\title{Photoluminescence signatures of
 negatively charged magnetoexcitons}
\author{James R. Chapman, Neil F. Johnson, 
V. Nikos Nicopoulos }
\address{ Department of Physics, Clarendon
 Laboratory, 
Oxford University,
Oxford OX1 3PU, England}
\maketitle

\begin{abstract}

The negatively charged quasi-two-dimensional
 exciton ($X^-$) is studied
 numerically in the presence of a
 uniform perpendicular B-field and various
 in-plane confinements.  The corresponding
 photoluminescence (PL)
 spectra, {\em including} shakeup
 processes, are calculated. These calculations
 provide a quantitative explanation of recent
 experimental
 PL spectra. 
 The photoluminescence signatures of
 charged excitons
 in quantum dots are also examined.
 Electron-hole pair distribution functions
 are calculated which show the 
structure of the charged exciton states
 (both ground and excited states).
 Increasing excitation reflects a
 competition between two
 effects: partial ionization of {\em either}
 one {\em or} two electrons. 
\end{abstract}

\newpage

\section{Introduction}

Excitons ($X$) are crucial to the understanding
 of optical spectra from
 quantum wells with a low electron density.
 If the electron density is
 increased the excitonic effects are destroyed
 by phase-space filling from
 the other electrons and a Fermi Edge Singularity
 (FES) may be seen. Recent
 experiments have shown that as the electron
 density is lowered, the FES
 evolves into a discrete line which is not an
 exciton but
 is instead attributed to
 the negatively charged exciton ($X^-=X+e^-$)
 ~\cite{Shields1}
~\cite{Finkelstein2}. The negatively
 charged exciton was first predicted by Lampert
 in analogy with the various
 ions of hydrogen ~\cite{Lampert1}. Further 
variational studies have
 been carried
 out in both three dimensions (3D)
~\cite{Munschy1} and
 two dimensions (2D) ~\cite{Stebe1}. It is found
 that the restriction to 2D
  increases the binding energy of the second electron
 to about ten times its 3D value. This should
 facilitate the
 observation of the $X^-$ and, consequently, most
 experimental studies are
 carried out in the quasi-2D environment of the
 quantum well. Negatively
 charged excitons were first observed in ${\rm
 CdTe/Cd_{1-x}Zn_{x}Te}$ quantum
 wells~\cite{Kheng1}~\cite{Kheng2} and were later
 seen 
in ${\rm GaAs/Ga_{1-x}Al_{x}As}$
 ~\cite{Shields1}~\cite{Finkelstein1}
~\cite{Finkelstein2}~\cite{Shields2}
~\cite{Buhmann1}~\cite{Shields3}. Most of the
 experimental spectra
 are obtained using polarized
 photoluminescence (PL) techniques. A constant
 B-field is often applied
 perpendicular to the well since this allows
 a charged exciton to form even
 at electron densities for which an FES would
 be seen in zero field
~\cite{Kheng2}.
Recently the low energy tails of the PL spectra
 have been examined
 and evidence of shakeup processes have been
 found~\cite{FinkelsteinSU}. 
These occur when
 the charged exciton undergoes electron-hole
 recombination leaving the remaining electron
 in an excited state. Positively charged excitons,
 consisting of an electron
 and two holes ($X^+=X+h^+$) have also been
 observed in p-tpye quantum
 wells~\cite{Finkelstein2}~\cite{ShieldsX+1}
~\cite{ShieldsX+2}.

Recent theoretical studies of the charged exciton
 complex consider
 ideal 2D systems ~\cite{Hawrylak1} and are
 often restricted to the
 lowest Landau level ~\cite{Palacios1}. In a
 previous paper we showed that
 the combined effects of (a) higher Landau
 level mixing and (b) the
 form factor emulating
 the finite quantum well width (i.e. quasi-2D
 system), are essential for
 understanding recent experimental
 $X^-$ spectra ~\cite{Chapman1}. 

In this paper accurate numerical solutions are
 presented 
for a quasi-2D negatively
 charged exciton ($X^-$) consisting of two
 electrons and a hole situated in 
an in-plane confinement potential and a perpendicular
 uniform B-field. The
 $X^-$ properties are studied over a wide
 range of experimentally accessible
 B-fields and in-plane confinements. We
 consider
 weak, intermediate and strong
 in-plane confinements, and use a
{\em range of initial angular momenta} for
 the charged exciton. We calculate
 the development with B-field of the
 following: 1) The photoluminescence spectrum
 2) the shakeup lines in the
 PL spectrum and 3) the electron-hole pair
 distribution functions.
For the case of a weak in-plane confinement
 we can compare our results to
 recent experiments in high mobility quantum wells. 
We find that 
theory and experiment match
 very closely in both the main
 spectrum (Sec.~\ref{PL})
 {\em and} the shakeup lines
 (Sec.~\ref{SU}). For
 stronger confinements our results should be
 compared to experiments carried
 out in quantum dots. To our knowledge no
 charged exciton species have yet
 been knowingly observed in quantum dots.
 We predict rich signatures 
 characterizing the $X^-$ in the quantum dot
 and hope that these results
 may stimulate
 further experimental study in such systems
 (Sec.~\ref{PL} and Sec.~\ref{SU}).
 Finally our study of the electron-hole pair
 distribution
 functions allows us understand the structure
 of the charged exciton species
 and how this relates to the features of our
 calculated spectra (Sec.~\ref{RPDF}). We also
 analyse the distribution functions of the
 excited $X^-$ states finding that
increasing excitation reflects a competition
 between two effects: partial 
ionization of {\em either} one {\em or}
 two electrons. 

\section{Model}
\label{Model}

The negatively charged exciton consists of
 two electrons and a hole
 interacting in a constant magnetic field
 according to the following
 Hamiltonian:

\begin{center}
\begin{eqnarray}
 {\mathcal{H}} &=& {1 \over 2m_{h}}\left[ 
{\bf p_{h}} - {e \over c}{\bf A}
({\bf r_{h}})\right]^{2} + {1 \over 2}m_{h} 
{\omega}_{h}^{2}{\bf r_{h}^{2}} 
+g_h \mu S^{z}_{h} \nonumber \\ &+& 
\sum_{i=1}^2 \left\{ {1 \over 2m_e}
 \left[ {\bf p_{e,i}} + {e \over c} {\bf A} 
({\bf r_{e,i}}) \right]^2 + {1 \over 2} 
m_e {\omega}^2_e {\bf r_{e,i}^2} +
 g_e \mu S_{e,i}^z - V_{eh}(|{\bf r_{e,i}}
 - {\bf r_h}|) \right\} \nonumber \\ 
&+& V_{ee}(|{\bf r_{e,1}} - {\bf r_{e,2}}|)
\end{eqnarray}
\end{center}
where $m_e = 0.067 m_0$ and $m_h = 
0.48 m_0$ are the masses of the
  conduction electrons and heavy holes respectively
 for GaAs. If the interactions $V_{ee}$, and
 $V_{eh}$ are set to zero and
  the symmetric gauge is chosen (i.e. ${\bf A}
({\bf r})=({\bf B \times r})/2$)
 the electron and hole single particle states
 are as follows:

\begin{center}
\begin{eqnarray}
\Psi^{e,h}_{n,M}({\bf r}) = A^{e,h}_{n,M} r^{|M|}
 {\rm e}^{iM \theta} L^{|M|}_n
\left[ {r^2 \over 2l^2_{e,h}} \right] {\rm exp}
 \left[{-r^2 \over
 4l^2_{e,h}}\right]
\end{eqnarray}
\end{center}
with single particle energies
\begin{center}
\begin{eqnarray}
E^0_e=\hbar {\overline \omega_e} \left[ n + {1
 \over 2} + {|M| \over 2} \right]
+ {\hbar \omega^c_e \over 2} M \\
E^0_h=\hbar {\overline \omega_h} \left[ n + {1
 \over 2} + {|M| \over 2} \right]
- {\hbar \omega^c_h \over 2} M 
\end{eqnarray}
\end{center}
$A^{e,h}_{n,M}$ is a normalisation, $L^{|M|}_n$
 are associated Laguerre
 polynomials, $\hbar \omega_e$ and $\hbar 
\omega_h$ are the parabolic
 confinements for the electron and the hole
 respectively, and $\omega^c_{e,h}
 = eB/m_{e,h}$ is the electron/hole cyclotron
 frequency. Also ${\overline
 \omega_{e,h}} =((\omega^c_{e,h})^2 + 4
 \omega^2_{e,h})$. The single particle
 wavefunctions have characteristic length scales
 $l^2_{e,h} = \hbar /m_{e,h}
{\overline \omega_{e,h}}$. In this symmetric
 gauge the single particle
 wavefunctions have two quantum numbers, n and M.
 In the limit of weak
 confinement $(i.e. (\omega^c_{e,h})^2 \gg
 4(\omega_{e,h})^2)$
 the magnetic field
 dominates
 and the single particle states form Landau
 levels. For electrons, states with
 $M_e \le 0$ and n=0 are degenerate forming 
the lowest Landau level;
 raising n or making $M_e$ positive raises 
the electron energy 
and puts it in a higher
 Landau level. For the holes the positive
 $M_h$ states with n=0 
 form the lowest hole Landau level.

 In the case of strong confinement
 $(i.e. (\omega^c_{e,h})^2 \ll 4(\omega_{e,h})^2)$
 the single particle states
 are essentially oscillator states of the
 parabolic confining potential,
 where ${\overline \omega_{e,h}} \rightarrow
 2 \omega_{e,h}$.
 For a given n, states with the same $|M|$
 are hence degenerate. Raising either n or $|M|$
 raises the energy by an integral
 number of $\hbar \omega_{e,h}$.
 For both electrons and holes, the states have
 definite angular momentum,
 $\hbar M_{e,h}$, arising from the rotational
 invariance of the system. The
 experimental data referred to in this paper 
is all for weak confinements.
 For simplicity, therefore, we refer to Landau
 levels
 throughout this paper noting that the
 Landau level is no-longer degenerate when the
 confinement is strong. The
 position of the particle in a given single
 particle state is closely
 related to the angular momentum of the state, M.
 For particles in the
 lowest Landau level (i.e. n=0):
\begin{eqnarray}
\left<r^2\right>=2l^2\left(|M|+1\right).
\label{<r^2>}
\end{eqnarray}

This paper considers three confinements: weak 
$(i.e. (\omega^c_{e,h})^2
 \gg 4(\omega_{e,h})^2)$, intermediate $(i.e.
 (\omega^c_{e,h})^2 \sim
 4(\omega_{e,h})^2)$ and strong $(i.e.
 (\omega^c_{e,h})^2
 \ll 4(\omega_{e,h})^2)$.
 The cyclotron energies are $\hbar \omega^c_e
 = 17.2 {\rm meV}$ and $\hbar
 \omega^c_h = {\rm 2.4meV}$ at B=10 T. The
 confinements are chosen to be
 $\hbar \omega_e =$1.75 meV in the weak
 confinement limit,  $\hbar \omega_e
 =$5.58 meV in the intermediate confinement 
limit and  $\hbar \omega_e
 =$17.6 meV in the strongly confined limit.
 For ease of comparison 
we chose $\omega_h$ such that
 $\omega_e/\omega_h = m_h/m_e$, thereby ensuring
 that the electron and hole
 wavefunctions have equal characteristic lengths,
 $l^2_{e,h}=\hbar/m_{e,h}
 {\overline \omega_{e,h}} = l^2$. In the case of
 weak confinement the 
magnetic field dominates and we can compare our
 results to those obtained 
from high mobility quantum wells. For intermediate
 and strong confinements 
the charged exciton is effectively in a quantum dot.
 In the
 case of intermediate confinement, the B-field
 dominates at reasonably large
 B while
 the confinement dominates at low B. The crossover
 occurs when $\omega^c_{e,h}
 \sim 2\omega_{e,h}$ i.e. when B$\sim$6.5T.
 As expected we find interesting
 changes of behaviour around the crossover field
 strength.

The interaction potentials are Coulombic. 
In order to obtain a
 tractable model it is assumed that the single
 particle wavefunctions
 separate: ${\Psi}_e = {\phi}_e(z_e)
{\psi}_e({\bf r_e})$
 and ${\Psi}_h = 
{\phi}_h(z_h){\psi}_h({\bf r_h})$.  A specific form
 for ${\phi}_{e,h}$ is chosen,
 thereby freezing out the z-motion and yielding
 a quasi-2D
 model. There are two different limits we can
 choose for ${\phi}_{e,h}$.
 In one limit the electrons and holes occupy
 the
 same plane but have their charge distributions
 smeared in a rod-like
 distribution along the perpendicular direction
 (rod geometry).
 In the other limit the electrons and holes
 exist on separate
 planes separated by a
 distance d (biplanar geometry), their charge
 distributions being unsmeared.
 The biplanar
 geometry is the relevant limit of systems
 such as a quantum well in a polarizing
 electric field which separates the electrons
 and holes. The rod geometry is
 suitable for quantum wells in which the electron
 and hole distributions are
 not separated in the direction perpendicular to
 the well. We showed in an
 earlier publication ~\cite{Chapman1} that the
 biplanar geometry doesn't
 exhibit a PL signature of charged excitons
 for $d > l$. This
 lack of $X^-$ signature has been
 confirmed experimentally ~\cite{ShieldsEfield} for
 a quantum well in a perpendicular
 electric field. Consequently we concentrate
 on
 the rod geometry in this paper, for which:
\begin{center}
\begin{eqnarray}
|{\phi}_{e,h}(z_{e,h})|^2 = \left\{
 \begin{array}{ll}
                          {1 \over L} 
 & \mbox{if ${-L \over 2} 
< z_{e,h} < {L \over 2}$}  \\
                          0  & 
 \mbox{otherwise}
                        \end{array}
                       \right. 
\end{eqnarray}
\end{center}

In this work we use a rod length $L=3l$.
 At B=10 T, $L=240$\AA \ for
 weak confinement, $L=220$\AA \ for
 intermediate confinement and $L=170$\AA
 \ for strong confinement.

Before proceeding with any calculation of
 charged exciton states it is
 important to identify the symmetries of
 the full Hamiltonian as these lead
 to well-defined quantum numbers, thereby
 allowing us to work within a
 restricted Hilbert space. For the charged
 exciton the Hamiltonian in the
 symmetric gauge is rotationally invariant
 and also conserves the total spin.
 The good quantum numbers for the charged
 exciton are therefore
 the total orbital angular momentum $\hbar
 {\rm M}$, the total
 spin $\hbar {\rm S(S+1)}$ and the total
 z-component of the spin, $\hbar 
{\rm S_z}$. Spin
 conservation, together with the fact that
 the wavefunction must be 
antisymmetric under interchange of the two
 electrons, allows us to label the
 charged exciton wavefunctions as either
 triplet or singlet. The singlet
 states have S=0 and ${\rm S_z=0}$ with an
 orbital wavefunction which is
 symmetric under electron interchange. The
 triplet states have S=1 and
 ${\rm S_z = \pm 1\ or \ 0}$ with an orbital
 wavefunction which is
 antisymmetric to electron interchange.
 Thus the charged exciton
 wavefunctions have the form:
\begin{eqnarray}
\Psi(M,S=0,S_z=0)= \sum_{n_1 M_1 \ge n_2 M_2}
 &C&^S_{n_1 M_1, n_2 M_2, n_h M_h}
\phi^S_{n_1 M_1, n_2 M_2, n_h M_h}(M)\nonumber
 \\  &\chi&(S=0, S_z=0)
\nonumber \\
\Psi(M,S=1,S_z=\pm1,0)= \sum_{n_1 M_1 >
 n_2 M_2} 
&C&^T_{n_1 M_1, n_2 M_2, n_h M_h}
\phi^T_{n_1 M_1, n_2 M_2, n_h M_h}(M)\nonumber
 \\ &\chi&(S=1, S_z=\pm1,0)
\end{eqnarray}
where
\begin{eqnarray}
\phi^S_{n_1 M_1, n_2 M_2, n_h M_h}(M)&=&{1
 \over \surd 2} 
\left\{ \phi_{n_1 M_1}
({\bf r_1})\phi_{n_2 M_2}({\bf r_2})+\phi_{n_1
 M_1}
({\bf r_2})\phi_{n_2 M_2}({\bf r_1}) \right\}
 \phi_{n_h M_h}({\bf r_h})
\nonumber \\
\phi^T_{n_1 M_1, n_2 M_2, n_h M_h}(M)&=&{1
 \over \surd 2}
 \left\{ \phi_{n_1 M_1}
({\bf r_1})\phi_{n_2 M_2}({\bf r_2})-\phi_{n_1
 M_1}
({\bf r_2})\phi_{n_2 M_2}({\bf r_1}) \right\}
 \phi_{n_h M_h}({\bf r_h})
\end{eqnarray}
Here $\chi(S=0,S_z=0)$, $\chi(S=1,s_z=\pm1,0)$
 are 
the usual singlet and triplet
 spin states respectively. The total angular
 momentum $M= M_1 + M_2 + M_h$.
 The restriction $n_1M_1 \ge n_2M_2$ is required
 to prevent overcounting
 of states.
In the particular limit that the basis is
 restricted to the
 lowest Landau level, the interaction
  $V_{eh}=V_{ee}$ and the in-plane confinement
 is zero,
 there is a well documented hidden symmetry
~\cite{MacDonald1}~\cite{Apalkov1}.
 This forces the photoluminescence energy from
 a triplet charged exciton to
 be indentical to that of the lowest exciton line. 
In our system this symmetry is broken
 by the in-plane confinement and by Landau level
 mixing,
 the latter being crucial 
in the case of weak confinement~\cite{Chapman1}.
 This breaking of the
 hidden symmetry is vital if any signature of
 the triplet charged exciton is
 to be seen in photoluminescence.

 The effects of the B-field and the Coulomb
 interaction compete: from a perturbation theory
 perspective the 
expansion parameters are given by ${a_{e,h} 
/ l}$ where $a_{e,h} = \epsilon
 \hbar^2/m_{e,h}e^2$ is the Bohr radius for 
the electron and hole respectively.
 Taking $B \sim 8$T and $\epsilon_r = 12.53$
 yields (GaAs) ${a_h / l} =0.2$ 
and ${a_e / l} = 1.1$, thereby rendering any such 
perturbation approach unreliable. 
For this reason we diagonalize the charged
 exciton Hamiltonian 
exactly numerically. We perform the
 diagonalizations in regions of the 
Hilbert space charaterized by quantum numbers
 for the charged exciton; i.e. 
 $M$, S and ${\rm S_z}$. The actual basis
 used has 5 hole Landau levels, 2
 electron Landau levels and 16 angular
 momentum states per Landau level.
 A larger number of hole Landau levels is used
 because of the small hole 
cyclotron energy. A finite number of angular
 momentum states is used since
 the confinement breaks the degeneracy of the
 Landau level.  Clearly as 
${\rm B} \rightarrow 0$ an increasingly
 large number of Landau levels should
 be used. We therefore only quote our calculations
 for B$\ge2$ T, the reliability
 of the answers increasing rapidly with B-field.
 From comparison with
 published experimental data given later in 
this paper and in Ref.
 ~\cite{Chapman1}
 the authors believe this basis to be sufficient
 for capturing the essential
 physics of the charged exciton. For the purpose of 
comparison, the exciton
 states are calculated using the same basis 
in a manner identical to
 that used for the charged exciton.
The reliability of the calculation has been
 verified  by reproducing known results
 as follows: 
(a) in the limit of zero confinement 
 the hidden symmetry result is recovered
 for B$\rightarrow \infty$, 
(b) the B=0 confined 2-electron results of Ref.
~\cite{Merkt1} are reproduced, 
(c) the main results found in a previous
 study on an ideal 2D charged exciton 
~\cite{Hawrylak1} are obtained.

\section{Photoluminescence Spectra}
\label{PL}

In this section we consider the photoluminescence
 (PL) spectra of charged and
 uncharged excitons. These spectra arise
 from the recombination of a valence
 hole with a conduction electron giving
 out a photon of energy $\hbar \omega$
 such that $\hbar \omega = E_i - E_f$ where
 $E_i$ and $E_f$ are the initial
 and final energies respectively. The
 recombination process conserves total
angular momentum. The emitted photon has angular
 momentum $M_\gamma = \pm 1$ 
 which is taken up by the atomic part of the
 Bloch wavefunction (i.e.
 transition between 
 s ans p orbitals) causing the angular momentum of
 the envelope function to remain unchanged
 under recombination. Throughout
 this paper
 we are only dealing with  envelope
 wavefunctions (i.e. we neglect bulk
${\bf K.p}$ bandmixing effects) hence we
 have the selection rule
 $\Delta M=0$. Consequently, only excitons
 with
 $M$=0 can undergo recombination and
 contribute to the PL spectrum ($X(M=0)
 \rightarrow \gamma$). Now consider recombination
 from charged excitons. 
Angular momentum must still be conserved, but
 on recombination an electron
 is left in the final state. This electron can
 have any angular momentum, 
thereby  allowing charged excitons with total
 angular momentum, M$\neq 0$
 also to contribute
 to the PL spectrum via $X^-(M) \rightarrow
 \gamma + e^-(M)$. The final
 electron can be in any Landau level provided
 it has angular momentum $M$.
 The higher the energy of the final state
 electron, the lower the PL photon
 energy. In this section we always consider
 electrons to be in the lowest
 Landau level in the final state. Electrons
 left in excited states (i.e.
 higher 
Landau levels) are considered later in this
 paper when we deal with shakeup
 processes. 

We consider the recombination to involve
 the $\pm {3 \over 2}$
 heavy holes; this is consistent with the
 relevant experimental work~\cite{Shields2}.
 These holes combine with
 $\pm {1 \over 2}$ electrons
 giving two circularly polarised lines,
 $\sigma^+$ and $\sigma^-$, which
 are Zeeman split by the magnetic field.
 The spin of the remaining electron
 from the charged exciton has no effect
 on this splitting as its Zeeman
 energy makes the same contribution to
 the energies of the initial and final
 states. Experimentally the Zeeman splitting
 of the exciton PL lines is
 found to be less than that of the charged
 exciton lines~\cite{Shields2}.
 For clarity therefore the calculated PL lines
 presented here are situated at the
 average energies of the $\sigma^+$
 and $\sigma^-$ lines (i.e.
 taking $g_{e,h} \rightarrow 0$). 

As well as the PL energies, we are
 interested in the PL strengths.
 Ignoring complications due to population
 dynamics, we will take 
the measure of the PL strengths to be $|
\left< 0| \hat{L} |X \right>|^2$ and
  $|\left< e^-| \hat{L}
 |X^-\right>|^2$ for $X$ and $X^-$
 respectively where $\hat{L} =
 \int \hat{\psi_e}({\bf r}) \hat{\psi_h}
({\bf r})d^2 {\bf r}$. This leads
 to expressions for PL strengths as follows:

\begin{eqnarray}
\left| \left<n_0 M_0| \hat{L} | S=1,S_z
 = \pm 1 \right> \right|^2 &=& 
\left| F^T \right|^2 \nonumber  \\
\left| \left<n_0 M_0| \hat{L} | S=1,S_z
 = 0 \right> \right|^2 &=& 
{1 \over 2}\left| F^T \right|^2 \nonumber  \\
\left| \left<n_0 M_0| \hat{L} | S=0,S_z
 = 0 \right> \right|^2 &=& 
\left|{1 \over \surd 2} F^S + C^{S}_{n_0
 M_0 , n_0 M_0, n_0 -M_0} \right|^2
  \nonumber \\
\left| \left<0| \hat{L} |X \right>
 \right|^2 &=& 
\left| F^X \right|^2	
\end{eqnarray}

where

\begin{eqnarray}
F^{T/S} &=& \sum_{n_0 M_0 >n -M}
 C^{T/S}_{n_0 M_0 , n -M , n M} -/+
 \sum_{n_0 M_0
 < n -M} C^{T/S}_{n -M , n_0 M_0 ,
 n M}\nonumber \\
F^X &=& \sum_{n,M} C^X_{n -M ,
 n M}\nonumber
\end{eqnarray}
Here $S$ is the total electron spin,
 $S_z$ is the $z$-component of total 
electron spin and $C^{S,T}$ are the expansion
 coefficients of the singlet/triplet
 charged exciton. The $C^X$ are the
 expansion coefficients of the exciton.
 The electron remaining from the
 charged exciton recombination is in
 the state with quantum numbers 
$\{ n_0,M_0\}$. It is difficult to
 predict which of the triplet (S=1)
 states contributes most to the PL spectrum.
 Therefore we use $|F^T|^2$
 as a measure of the PL strength
 of the triplet line.

We now turn to the photoluminescence  
(PL) spectra of charged and neutral
 excitons in a weak in-plane confinement.
 At all B-fields examined (i.e. B=$2 
\rightarrow 12$ Tesla) we find that the 
ground state of the neutral 
exciton (X) has 
 total angular
 momentum $M=0$, the ground state of the
 singlet charged exciton ($X^-_s$) has
  $M=0$ and the ground state of the triplet
 charged exciton ($X^-_t$)
 has $M=-1$. These all contribute to the PL
 spectrum. Other
 charged exciton states with M$\leq 0$ can
 also luminesce
 via $X^-(-{\rm M}) \rightarrow \gamma +
 e^-(-{\rm M})$; the electron 
in the final state
 is left in the lowest Landau level. Since
 these states are very close in 
energy to the absolute $X^-_s$ and $X^-_t$
 ground states and their PL lines
 are also similar, they must be included
 in a calculation of the PL spectrum.
 Increasing $|M|$ increases the energy of
 the $X^-$ state, so
 we only need to consider those states for
 which $-3\leq$M$\leq0$
 in our calculations. 

In a weakly confined system it makes sense 
to discuss whether the $X^-$
 species are stable to the decomposition
 $X^- \rightarrow X + e^-$.
A charged exciton will decompose if its 
initial energy is greater than that
 of the final state consisting of an exciton
 in its lowest energy state and
 a free electron in the lowest Landau level.
 In a truly unconfined system,
 the angular momentum of the final state
 electron is irrelevant due to the
 degeneracy of the Landau level. In our
 calculations however, the electron 
single particle energies do vary weakly
 with ${\rm M_e}$ thus we assume
 angular momentum conservation in order
 to take account of this; i.e. $X^-(M)
 \rightarrow X(0) + e^-(M)$ where the $X(0)$
 state is the ground state
 of the magnetoexciton. 
 This decomposition
 is closely related to the PL recombination
 process for $X^{-}$. Since $X(0)$ 
can recombine directly ($X \rightarrow 
\gamma$) it is clear that an $X^-$ 
state will be stable to this decomposition
 provided its PL energy is lower
 than that from the bare exciton. 

Figure 1 shows the development of the
 spectrum with B-field for a rod length
 $L=3l$ (240\AA \ at B=10T) and a weak
 in-plane confinement.
 To facilitate computation, the rod length
 is allowed to scale with the characteristic
 length $l$. Recent work on the
 interface exciton indeed suggests that
 some reduction of $L$
 with B is expected
 ~\cite{Viet}. 
 The spectra are robust to changes in
 $L$ and  hence Fig.1 retains
  the experimental features simulated 
in our earlier paper
 ~\cite{Chapman1} using fixed $L=220$\AA.
 Reference~\cite{Chapman1}
 only fully considered PL spectra from
 $X^-$ species with $M=0$.
 In contrast Fig.1 includes optical
 recombination from the lowest energy
 singlet and
 triplet states corresponding to
  $M=-3,-2,-1$ and $0$.
 The $X^-_s$, $X^-_t$ and exciton ($X$)
 PL spectra are shown separately. A Gaussian
 broadening of $0.3$meV FWHM
 has been 
introduced.

The spectra in Figure 1 agree well with those
 observed experimentally 
in quantum wells of
 width 200 -- 300\AA ~\cite{Shields2}.
 The PL energies
 increase with B in a similar way to
 the experimental data~\cite{Shields2};
in addition
 the theoretical spectrum consists of
 a predominately $X^-_s$ peak
 about 1meV below the exciton line at 8T 
(as compared to the experimentally observed
 value of 1.7meV at 8T).
 At low B Fig.1 shows an $X^-_t$ peak
 above the exciton line. This is due
 to recombination from the $M=0$ triplet
 charged exciton. This will
most likely not be
 observed experimentally since its position
 indicates that it is unstable to 
decompostion into an exciton and
 electron (as discussed earlier) and as such
 is not expected to 
form. However, on increasing B, this
 $X^-_t$ line can be seen to move through
 the  exciton line, thus becoming stable
 to decomposition and allowing it 
to be observed. This occurs at around
 6 Tesla. The appearance of an $X^-_t$
 line near the exciton ($X$) line at
 finite B
 is also a feature of the experimental
 spectra in which it occurs at around
 2.5T ~\cite{Shields2}.
 The lower energy contribution to
  the triplet spectra in Fig.1 originates
 from the $X^-_t$ states with M$<0$.
 As B is increased these clearly lose
 strength to the M=0 triplet line.
 This is a precursor to the hidden symmetry
 mentioned earlier; as B increases
 the low energy states are forced to
 have more weight in the lowest
 electron and hole Landau
 levels. Apart from the M$<0$ triplet lines,
 the PL strength of the
 lines generally increases with B. This
 is much
 more pronounced for the exciton ($X$)
 and $X^-_t$ than for $X^-_s$.
 This effect
 is observed experimentally ~\cite{Shields2}
 and agrees quite 
closely with Fig.1. The splitting between the 
exciton ($X$) and $X^-_s$ lines in Fig.1
 increases with B in a manner similar to that
 observed experimentally and 
 is more pronounced than that found in our
 previous study which used a
 fixed rod length ~\cite{Chapman1}. This
 suggests that the observed
 increased
 binding of the $X^-_s$ with respect to the
 exciton is to a certain extent
 due to the scaling with magnetic length of
 the extent of the exciton and 
charged exciton wavefunctions perpendicular
 to the well.
We emphasize that the spectra presented in
 Fig.1 contain recombination
 from charged excitons with initial angular
 momenta, $M=-3,-2,-1$, and 0. The
 calculated spectra clearly have all the
 qualitative, and many of the
 quantitative features of the experimental
 spectra. Thus charged excitons
 with a range of initial angular momenta
 could be involved in the
 experimental systems reported to date.

We now consider the PL signatures of charged
 and uncharged excitons in an 
intermediate in-plane confinement potential
 ($\omega_e \sim \omega_c$). The 
ground states  are found to have angular
 momentum $M=0$ for the singlet charged
 exciton, $M=-1$ for the triplet and $M=0$
 for the exciton. This is identical to
 the weakly confined case. Again we examine
 the PL spectrum
 from charged excitons with
 M$\leq0$ as these undergo recombination
 leaving the remaining electron in a
 state of M$\leq0$ which becomes part of
 the lowest Landau level at high
 enough B-field. 

Fig.2 shows the calculated PL spectra for
 the lowest energy singlet and
 triplet $X^-$ states with $-3 \leq$M$\leq 0$
 together with the PL line from
 the ground state of the exciton (M=0). 
The exciton is included for
 comparison but would not be observable in
 the same quantum dot as a charged
 exciton since the parabolic confinement
 holds the electrons and holes 
together thus preventing complete ionization.
  The same rod length is used as in Fig.1. The
 in-plane spatial extent
 of
 the charged excitons ranges from
  300\AA \ to 400\AA \ at B=10T
 (discussed later) which is larger than
 the rod length, $L$ (220 \AA \ at B=10T)
 ensuring that the charged
 excitons retain the quasi-2D
 nature required by our model.
 These spectra contain interesting structure.
 The singlet ($X^-_s$)
 spectrum is
 always below the exciton ($X$) line and the
 triplet ($X^-_t$)
 spectrum has two parts, one 
higher than the exciton line which corresponds
 to the M=0 triplet line,
 and one at similar
 energies to the singlet spectrum which
 corresponds to the $M\ne0$
 triplet states. 
As B is increased the $X$--$X^-_s$ splitting
 increases 
slightly (1.2meV at 2T to
 1.5meV at 12 T) in a similar manner to that
 seen in the weakly confined 
case. We also note that the major group of
 lines show a
 diamagnetic shift upwards in energy
 with increasing B.
 More interesting is the development of the
 singlet and triplet 
spectra with B. We shall first consider 
the singlet spectra. At low B the 
only strong line arises from the singlet
 ground state ($M=0$). 
The $M=-1,-2,-3$
 lines are arranged below this decreasing
 in both PL intensity and energy 
as $M$ decreases. As B is increased these 
lower $M$ lines increase their PL 
strengths and energies moving through the
 $M=0$
 singlet line around B=7T. At high
 B the lower $M$ singlet lines are clearly
 arranged at higher PL energies than
 the $M=0$ line with PL strengths 
sufficiently high to allow them to be 
observed. The ordering with increasing PL
 energy is $M=0,-1,-2,-3$ at high B.
 The triplet
 spectrum below the exciton line shows
 the same development with
 B-field. As B is increased the $M=-2,-3$
 triplet lines gain PL strength and
 move through the ground state ($M=-1$)
 triplet line around B=7T arranging
 themselves in the order $M=-1,-2,-3$ 
with increasing  PL energy at high B
. The interesting development of this 
structure with B is due to the 
competition between the effects of the
 B-field and the 
parabolic confinement mentioned in
 Sec.~\ref{Model}. This competition
 is very important
 in this case of intermediate confinement.
 At low B the parabolic in-plane 
confinement dominates; this removes the
 degeneracy of the Landau
 levels and vastly lowers the number of
 states available to the 
charged exciton with which to minimize
 its electrostatic energy.
In the lowest energy
classical configurations the hole has
 $M_h = 0$ while the electrons
 have $M_{e_1} \approx M_{e_2}$ such
 that $M_{e_1} +
 M_{e_2} = M$, where $M$ is the angular
 momentum of the charged exciton.
 This configuration
 has negligible overlap between electron
 and hole single particle 
wavefunctions giving the
 very low PL strength at low B for higher
 $|M|$ as observed in Fig.2. As B
 is increased the effect of the 
confinement begins to disappear
 as the Landau levels
 form. The degeneracy at a given $M_e$
 and $M_h$ now allows
 the electrostatic energy to
 be minimized with finite single particle
 electron-hole overlap
 giving the stronger PL
 strength. This competition between B-field
 and confinement also explains
 the changes in the PL energy orderings.
 When the confinement dominates 
(i.e.low B-field) the
 charged excitons with higher $|M|$ have
 to leave their extra electron
 in a higher $|M|$ (i.e. higher energy) 
state after recombination thus pushing
 the PL line to lower energies. This effect
 disappears as B increases 
and begins to dominate since
 the degeneracy of the Landau levels makes
 the energies of the final $M$ states
 increasingly similar.
It makes sense, therefore, that the
 cross-over
 between these effects occurs around
 6.5T, which is precisely where the
 expansion parameter,
 $2\omega_{e,h} / \omega^c_{e,h}=1$.

Finally we turn to the case of a strong 
in-plane confinement. Fig.3
 shows the PL spectra for a rod length 
$L=3l$ and the strong confinement 
potential. Again the lowest
 energy state of the singlet charged exciton
 has total angular momentum $M=0$,
 the
 triplet $M=-1$, and the exciton $M=0$.
 We consider
 here photoluminescence from 
the lowest energy singlet and triplet states 
with angular momenta $-3
 \leq M\leq 0$. The exciton PL from the
 ground state ($M=0$) is also shown for
 comparison. However, as with the case 
of intermediate confinement,
 decomposition (i.e. ionization) of the
 charged excitons
 is not possible here since the strong
 confinement keeps the number of particles 
in the quantum dot constant. We note that
 the spectral lines posess 
a diamagnetic
 shift upwards as B increases.
 The form of the spectrum is essentially
 unchanged over the range of B-fields 
shown. This arises because the confinement
 dominates at all B-fields shown.
 Thus we see a spectrum which is very
 similar to the low B-field spectra in
 Fig.2. The only lines with significant
 PL strength are the
 ground state singlet ($M=0$), the ground
 state triplet ($M=-1$) together with 
the lowest $M=0$ triplet state, and the
 $M=0$ exciton ground state. The PL 
strengths of the other lines are low due
 to the strong confinement as 
explained above for the intermediate 
confinement case. The $M=0$ triplet 
line lies far above the exciton line.
 The $M=0$ singlet, $M=-1$
 triplet and $M=0$
 exciton lines are grouped within 4meV 
of each other for all
 B-fields in Fig.3. The triplet
 $M=-1$ line is now the lowest PL line,
 then comes the singlet $M=0$
 followed by the
 exciton. It is only with this very strong
 confinement that the triplet line
 drops below the singlet in the PL spectra.

\section{Shakeup Processes}
\label{SU}

In the previous section we considered 
recombination from charged excitons in
 which the electron in the final state
 was left in the lowest Landau level,
 i.e. $n_e = 0$ and $ M_e\leq 0$.
 We now consider recombinations 
in which the
 final electron is left in higher Landau 
levels. These are known as shakeup
 processes. We will concentrate on cases 
where the final electron is left in
 the first excited Landau level. These we
 denote as ${\rm SU_1}$ lines in the
 PL spectra. There are two ways in which
 an ${\rm SU_1}$ PL line may occur. In
 both cases the process is $X^-(M) 
\rightarrow \gamma + 
e^-(n_e,M_e=M)$; one case has
 $n_e =1,M_e\leq0$, the other involves
 $n_e=0,M_e=1$. 
We label the two processes
 ${\rm SU_1(n=1)}$ and ${\rm SU_1(n=0)}$
 respectively. These
 shakeup processes have significantly lower
 PL strengths than the 
recombinations dealt with in the 
previous section. They
 still obey the $\Delta M=0$
 selection rule, 
however the dipole matrix
 element is very small. In all Figures
 showing shakeup processes
 in this paper the strong PL spectrum 
is also shown for comparison.

Figures 4(a) and (b) show the ${\rm SU_1(n=1)}$
 and ${\rm SU_1(n=0)}$ PL lines
 respectively for the case of a weak
 in-plane confinement with rod
 length $L=3l$.
These spectra are very similar to recent
 experimental data for SU processes
 obtained in
 quantum wells ~\cite{FinkelsteinSU}.
 Consider Figure 4(a) which shows 
the ${\rm SU_1(n=1)}$ processes. The
 energies agree very well with the
 observed values (eg the splitting
 between the shakeup $X^-$
 lines and the non-shakeup $X^-$ lines
 is 7.4meV at 4T in Fig.4(a)
 and 6meV experimentally~\cite{FinkelsteinSU}).
 As B 
increases the SU lines drop
 almost linearly. Given the weak
 confinement we expect the difference 
 between the non-shakeup $X^-$
 line and
 the shakeup line to be $\hbar \omega^c_e$,
 the cyclotron frequency associated
 with the remaining excited electron.
 This is indeed what we find; the ${\rm
 SU_1(n=1)}$ line drops by 1.3meV/T while
 the nonshakeup $X^-$ rises by
 about 0.4meV/T, the difference between
 the two lines being 1.7meV/T
 corresponding to $\hbar \omega^c_e$ as expected.
 Experimentally the ${\rm SU_1}$ line
 is found to vary linearly with B,
 however the difference between it and
 the $X^-$ is slightly less than
 $\hbar \omega^c_e$ which has been
 tentatively attributed to an increase in the
 electron mass due to penetration of the
 AlGaAs layer ~\cite{FinkelsteinSU}.
 Looking in more detail at the calculated spectrum
 we see that the singlet ${\rm SU_1(n=1)}$
 lines are very weak in comparison
 to the triplet ${\rm SU_1(n=1)}$ lines.
 The latter dominate the
 ${\rm SU_1(n=1)}$ spectrum at low B-field.
 At B=2T the SU spectrum 
is about 100 times weaker than the 
strong non-shakeup $X^-$ line,
 this compares very
 well with the observed intensity ratio from
 Ref.~\cite{FinkelsteinSU}. As B increases
 the main triplet SU line drops rapidly
 in PL strength thus rendering the
 ${\rm SU_1(n=1)}$ unobservable at higher
 B. This is again found
 experimentally. The main triplet line 
consists predominately of
 $M=-1,-3$ lines, the $M=0,-2$ lines being
 similar in strength to the singlet
 lines. The large drop in shakeup PL
 strength with B is due to the
 increased energy
 difference between Landau levels causing
 the initial charged exciton states
 to have less contribution from both
 electron and hole higher Landau levels. 

 Fig.4(b) shows the ${\rm SU_1(n=0)}$
 processes. Again the
 energies agree well with observed
 values~\cite{FinkelsteinSU}.
 The SU spectrum is in fact almost
 identical to that of Fig.4(a) except
 that now the singlet line dominates
 and the triplet line is very weak.
 Again at B=2T the SU line is about 100
 times smaller than the strong non-shakeup
 $X^-$ line. The gradient of the
 ${\rm SU_1(n=0)}$ line is 1.3meV/T,
 the same as that in Fig.4(a).
  The ${\rm SU_1(n=0)}$ line
  loses strength rapidly
 with increasing B-field. This is again
 due to the increased energy
 difference between Landau levels causing
 the charged exciton to
 have less contribution from both 
electron and hole 
higher Landau levels. While
 both ${\rm SU_1(n=1)}$ and ${\rm SU_1(n=0)}$
 processes match very well with
 the observed ${\rm SU_1}$ line, it is 
likely that
 the majority of the observed PL strength
 actually
 comes from the ${\rm SU_1(n=1)}$ process 
which is dominated by the $M=-1$
 triplet line. This is the lowest energy
 triplet state and has a weak PL
 strength for a recombination process 
which leaves the final electron in the 
lowest Landau 
level thus suggesting that there will be a large
 population to undergo recombination via a shakeup
 process.

Figures 5(a) and (b) show the ${\rm SU_1(n=1)}$
 and ${\rm SU_1(n=0)}$ lines
 respectively for the case of an intermediate
 in-plane confinement with rod 
length $L=3l$. Again the non-shakeup lines 
are included for comparison. In
 both Fig.5(a) and (b) the PL strengths of
 the ${\rm SU_1}$ lines are
 significantly lower than the non-shakeup
 spectrum. First we consider the
 ${\rm SU_1(n=1)}$ lines in Fig.5(a). The
 overall
 ${\rm SU_1}$ spectrum is weaker than for 
the case of weak confinement (cf.
 Fig.4(a)). The triplet ${\rm SU_1}$ lines
 dominate over the singlet lines
 and there is no longer  a linear drop of PL
 energy with increasing B-field.
 Also the splitting between the shakeup 
$X^-$ lines 
and the non-shakeup $X^-$ lines does not
 change linearly with B.
 This is due to the confinement which
 prevents $\hbar
 {\overline \omega_{e,h}}$ from varying 
linearly with B. 
 The competition between confinement and
 B-field was important in the
 non-shakeup processes of Sec.~\ref{PL}, 
spreading the PL energies
 of the lines and affecting
 their strengths. Here we also see an effect
 on the PL energies of the 
${\rm SU_1(n=1)}$ lines. At low B-field, where
 the confinement dominates, the
 triplet ${\rm SU }$ PL lines are separated,
 the $M=-3$ state having the
 lowest PL energy then $M=-2,-1$ and finally
 $M=0$ with the highest energy. 
The energy separation is due to the confinement
 which raises the energy of the
 $-M$ final electron state thus lowering
 the PL energy. This effect
 is removed
 as Landau levels form with higher B-field
 (B$>$6.5T). The M=0
 triplet ${\rm SU_1}$ line always appears
 at higher energy because its
 initial state is higher in energy.

We now consider the ${\rm SU_1(n=0)}$ processes
 of Fig.5(b). The singlet
 spectrum clearly dominates the overall
 shakeup spectrum. It decreases
 linearly with increasing B and its intensity
 decreases with B also. The
 energy of the singlet ${\rm SU_1(n=0)}$ line
 is higher than the
  ${\rm SU_1(n=1)}$ lines of Fig.5(a). Finally
 we notice that the high PL 
 strength of the triplet ${\rm SU_1(n=0)}$ 
 line at B=2T is large, being
 in fact only ten
 times weaker than the non-shakeup lines. This
 occurs because the dominant
 effect of the
 confinement at low B-field is to force
 the $\pm M_{e,h}$ states to have similar
 energies, thereby causing the M=1 triplet
 charged exciton to configure itself more
 like the M$\leq0$ charged excitons and
 luminesce accordingly.

Figures 6(a) and (b) show the ${\rm SU_1(n=1)}$
 and ${\rm SU_1(n=0)}$ spectra
 respectively for the case of strong in-plane
 confinement, with a rod length
 L$=3l$. Again the non-shakeup spectra are shown
 for comparison. Considering
 the ${\rm SU_1(n=1)}$ processes in Fig.6(a),
 we see that the triplet
 lines dominate the singlet lines. The triplet
 spectrum is clearly split 
 into four sets each corresponding to a 
shakeup process from a
 charged exciton with a given $M$. The
 $M=-3$ has the lowest SU energy followed 
by $M=-2,-1$
 and finally $M=0$. This separation
 is due to the strong confinement (as in the
 low field regime of Fig.5(a))
  which raises the energy of the $-{\rm M_e}$
 final state electron
 thus lowering
 the PL energy. As B increases, 
the $-{\rm M_e}$ states start to
 converge in energy to form
 the Landau level; the spacing between the
 lines is consequently reduced. The
 ${\rm SU_1(n=1)}$ lines are very weak in
 this strong confinement limit.
 This is because the strong confinement 
forces the charged exciton
 to have very little contribution from the
 $n\ne0$ states which now have
 much higher single particle energies. 

The ${\rm SU_1(n=0)}$ lines of Fig.6(b) are
 much higher in PL energy than
 those of Fig.6(a). The singlet line decreases
 linearly with B having a 
gradient of about 1meV/T. The triplet line 
is not linear, and for low fields
 is close to the non-shakeup spectrum in both
 energy and PL strength. As in
 the case of intermediate confinement at low B,
 this is due to the similarity
 in energy
 of $\pm M_{e,h}$ states, thereby
 causing the $M=1$ triplet charged exciton to
 configure itself in a similar manner to the
 $M=-1$ state and luminesce
 accordingly. We notice this effect dies away
 with increasing B. However,
 it clearly persists to higher B than in the
 case of intermediate confinement.

From Figs 4,5 and 6 we see that the
 ${\rm SU_1(n=1)}$
 spectra become increasingly
 disimilar from the ${\rm SU_1(n=0)}$ spectra 
as the in-plane confinement
 is increased. This occurs because the
 degeneracy of the Landau levels is lost
 as the confinement is increased.
 The single particle states
 instead form themselves into harmonic 
oscillator levels. Thus for
 all confinements the
 ${\rm SU_1(n=1)}$ processes leave the final
 electron in an excited state,
 whereas in the ${\rm SU_1(n=0)}$ processes,
 the final state electron is
 increasingly in the lowest M=1 oscillator
 state. As such the
 ${\rm SU_1(n=0)}$ process is more like
 a normal recombination process than a shakeup process.

\section{Electron-Hole Distribution Functions}
\label{RPDF}

Our exact diagonalization of the charged
 exciton Hamiltonian gives us the
 full expansion coefficients for the various
 charged exciton wavefunctions.
 We can use these to investigate the structure of 
the charged exciton states. In particular, we
 are interested in how the electrons are
 correlated with the valence
 hole. This is conveniently visualized using
 the electron-hole pair distribution
 function, $g({\bf r})$ ~\cite{Maksym1}:

\begin{eqnarray}
g({\bf r})= \int P({\bf r} + {\bf r_0},
{\bf r_0})d^2{\bf r_0}
\end{eqnarray}
where
\begin{eqnarray}
P({\bf r},{\bf r_0})=\left< \sum^{2}_{i=1}
 \delta({\bf r_{e_i}}-{\bf r})
\delta({\bf r_h}-{\bf r_0}) \right>
\end{eqnarray}
Thus $g({\bf r})$ can be interpreted
 as the electron density relative 
to the hole and $P({\bf r},{\bf r_0})$
 as the probability of finding an electron
 at ${\bf r}$ and a hole at
 ${\bf r_0}$. $P({\bf r},{\bf r_0})$ is
 written in terms of our previous
 singlet and triplet wavefunction expansion
 coefficients ($C^S$ and $C^T$)
 as follows:

\begin{eqnarray}
P^{S/T}({\bf r},{\bf r_0})=\sum_{n m} 
\left| A^{S/T}_{n m} \pm B^{S/T}_{n m}
 \right|^2
\end{eqnarray}
where S takes the positive sign, T takes 
the negative sign and $A$ and $B$ 
are given by:
\begin{eqnarray}
A^T_{n m}&=&\sum_{n_1 m_1 > n m} C^T_{n_1 m_1
, n m, n_h m_h} \Psi_{n_h m_h}
({\bf r_0}) \Psi_{n_1 m_1}({\bf r}) \nonumber \\
 B^T_{n m}&=&\sum_{n m > n_2 m_2}
 C^T_{n m, n_2 m_2, n_h m_h} \Psi_{n_h m_h}
({\bf r_0}) \Psi_{n_2 m_2}
({\bf r})\nonumber \\
A^S_{n m}&=&A^{T \rightarrow S}_{n m} + \sqrt{2}
 \sum C^S_{n m,n m,n_h m_h}
\Psi_{n_h m_h}
({\bf r_0}) \Psi_{n m}({\bf r}) \nonumber \\
B^S_{n m}&=&B^{T\rightarrow S}_{n m}
\end{eqnarray}

The pair distribution functions are radially
 symmetric so we are able to plot
 the radial pair distribution function (RPDF)
 given by
 $\rho(r)=2 \pi r g(|{\bf r}|)$. Fig.7
 shows how these RPDFs are used to determine
 the structure of the charged
 exciton. If both electrons are equally
 correlated with the hole (Fig.7(a))
 then we expect $\rho(r)$ to have a single 
peak as in Fig.7(b). If however
 one of the electrons is closer to the hole
 than the other, as in Fig.7(c),
 then we expect to see two peaks corresponding
 to the inner and outer
 electron rings as in Fig.7(d). Charged
 excitons with this doubly peaked 
RPDF have a structure which is more like an
 exciton with an orbiting outer
 electron ($X+e^-$)

We now examine in more detail the
 electron-hole radial pair distribution
 functions
 for a charged exciton with rod length
 L$=3l$. Figure 8 shows how the 
electron-hole
 RPDF for the lowest energy singlet
 state (M=0) with a 
weak in-plane confinement varies with
 B-field. As B
 increases the electrons become more
 closely correlated with the hole.
As is generally the case, the
 overall form of the functions doesn't
 change; instead
 they just scale with the
 characteristic length $l$. In the case 
of weak confinement, this
 length is dominated by the magnetic 
length giving rise to a strong scaling
 with B. For stronger confinements the 
characteristic length becomes more
 dependent on the confining potential, 
thereby reducing
 the effects of changing
 B-field on the distribution functions.

Figures 9 and 10 demonstrate how the
 electron-hole RPDFs of
 the lowest energy singlet and triplet 
states for various $M$ depend on the
 confining potential. Figures 9(a) and
 (b) are for weak in-plane confinement,
 at  B=10T, with singlet and triplet
 charged excitons shown
 respectively. Figure 9(a) exhibits
 two different types of curve. 
The $M\leq 0$ curves
 are fairly narrow curves peaked at
 around 100 \AA \ indicating that
 both electrons are equally close to
 the hole. The $M>$0 curves are broader,
 peaking at between 200 and 300\AA. 
The $M=3$ curve even shows a shoulder at
 higher electron-hole separation indicating
 an $X+e^-$ structure. Figure 9(b) shows the
 corresponding triplet curves. Again 
there are two types of curve.
 The triplet $M<0$ electron-hole pair
 distributions are
 narrow and peaked at about 150\AA, 
whereas the $M\geq0$
 curves are much broader.
 As $M$ increases from 0 to 3 a 
distinct shoulder develops at higher
 electron-hole
 separation indicating the $X+e^-$ structure.
 In general, the triplet curves
 broaden more quickly than the singlet
 curves with increasing $M$. Of 
particular interest is the $M=0$ triplet
 electron-hole
 RPDF. This shows a broadening that
 the corresponding singlet curve does
 not. The reason for this arises from
 the symmetries of the wavefunctions. 
The singlet charged exciton has 
electrons which are symmetric under 
interchange of their spatial
 wavefunctions. Thus the $M=0$ singlet
 charged exciton can have both electrons
 and the hole with the same
 angular momentum $M_{e,h}=0$. This gives rise
 to the sharply peaked singlet $M=0$
 curve in Fig.9(b).
 The triplet charged exciton
 however has a spatial wavefunction which 
is antisymmetric under electron
 interchange. This implies that the lowest
 energy $M=0$ triplet state
 cannot be built with both electrons and
 the hole in states with
 $M_{e,h}=0$. Instead the state will have
 one electron with
 $M_e=0$, the other with $M_e=-M$ and the hole
 with $M_h=M$. A state with $M_h<M$ would
 force the electron with $M_e=0$ to change
 to a state of higher ${\rm M_e}$,
 thus placing it in a higher Landau
 level; this is 
energetically unfavourable. A state
 with $M_h > M$ would
 force the hole into a higher $|M|$ state
 than either electron; 
 the electrons would hence come closer
 in $M_e$. This situation is clearly 
electrostatically unfavourable given the
 relation between $|M|$
 and $<r^2>$ provided by Eqn.~\ref{<r^2>}.
 Thus from symmetry and energy 
 considerations, the lowest energy $M=0$
 triplet charged exciton should have one
 electron closer
 to the hole than the other. This gives 
rise to the broad electron-hole
 RPDF seen in Fig.9(b).
 It also partially explains the fact that
 the triplet ground state has
 angular momentum $M=-1$, whereas the
 singlet ground state has $M=0$. The 
 electron-hole RPDFs at B=10T in the case 
of strong confinement
 are shown in Fig.10. As with Fig.9,
 only the lowest energy states at each
 M are considered.
 Figure 10(a) shows the electron-hole RPDF for the singlet 
charged exciton. The $M=0$
 case is still the most sharply peaked.
 The other RPDFs all have similar peak
 heights and widths.
 This is a direct result of the strong
 confinement which lifts
 the degeneracy of the Landau levels
 instead giving instead single
 particle states corresponding to a
 harmonic oscillator. Now the
 degeneracy is between single particle 
states with the same $|M|$. This
 forces the $-M$ charged exciton states to
 behave very like those with $+M$.
 This can be seen in the RPDFs of Fig.10(a) 
 where the $M=-1$ curve resembles the
 $M=+1$ curve etc. Comparing Figs.10(a)
 and 9(a) we see that the distribution
 functions peak at similar electron-hole 
separations in both cases. The high
 confinement is conspicuous however in removing
 the shoulders corresponding to high
 electron-hole separation
 from the distribution functions;
 consequently all
these lowest energy states have both
 their electrons equally well
 correlated with the hole. Figure 
10(b) shows the corresponding
 electron-hole RPDFs for
 the triplet charged excitons. This
 shows strong similarities with the singlet
 case. The main distinction is the
 sharply peaked triplet $M=-1$  RPDF which
 is narrower than the others. This
 corresponds to the $M=0$ RPDF in the singlet
 case (Fig.10(a)). It comes as no
 surprise therefore that
 the $M=0$ state is the lowest
 energy singlet state whilst the lowest
 energy triplet state has $M=-1$. As
 in the weakly confined case this difference
 in the angular momenta of the
 lowest energy singlet and triplet states 
is due to the symmetries of the
 singlet and triplet spatial wavefunctions
 under electron interchange. Again
 we see the triplet states showing evidence
 of underlying oscillator single
 particle states since the $+M$ and $-M$
 states have become very similar.

The electron-hole RPDFs 
for an intermediate in-plane
 confinement behave as the strong or weak
  confinement cases depending
 whether the B-field or the confinement
 is dominant. Earlier we saw that
 the cross-over between the dominance of
 the confinement or the
 B-field occurs at about B=6.5T for this
 intermediate
 confinement. Below B=6.5T the electron-hole
 RPDFs are similar to those in Fig.10. Above
 B=6.5T the 
electron-hole RPDFs are similar to those in Fig.9.

We now turn to the excited states. As an
 illustration we look at the
 excited states with the same $M$ as the
 lowest energy singlet ($M=0$) and
 triplet ($M=-1$) charged excitons. The 
electron-hole
 RPDFs for the first five states of
 these systems are shown in Figs.11(a) and (b).
 A charged exciton can become excited
 in essentially two different ways. First 
it can start to ionize itself
 by moving the second electron increasingly
 further away from the electron
 and hole which are left in the centre.
 Second it can keep both electrons equally
 close to the hole, but alter their correlations
 with the hole. In the first
 case we should see two peaks forming in the
 electron-hole distribution
 function. In the second the distribution
 function should have a single peak,
 but the behaviour near the origin and the peak
 position should be modified
. With this in mind we consider Fig.11. First
 we consider the singlet
 case (Fig.11(a)). The lowest four energy states
 clearly show the 
 electron-hole RPDF
 becoming broader as the energy increases,
 until eventually at level 4 there is a definite
 high electron-hole
 separation shoulder indicating a partial
 separation into $X+e^-$. The next
 energy level up sees both electrons equally
 correlated with the hole, the
  RPDF vanishing more quickly at the origin
 and its peak position having moved
 to higher separations. This corresponds
 to the second type of excitation.
 The triplet energy levels at $M=-1$ (Fig.11(b))
 show a very similar
 behaviour. The electron-hole RPDFs broaden
 as the energy level is
 increased. A higher separation shoulder
 can be just distinguished
 in the RPDF for
 level 5.

Similar effects are seen in the excited 
states at different M and different
 confinements. The clearest case for showing 
the two types of excitation
 process is a strongly confined charged
 exciton with $M=-$2 at B=10T. This
 is shown in Fig.12. Here the first three 
levels show  
 broad 
 electron-hole RPDFs. Level 4 has two
 distinct peaks
 in its RPDF arising from two
 different electron rings around the hole.
 Level 5 then has an electron-hole
 RPDF which
 has one peak moved to slightly higher e-h
 separation, and clearly has a
 different behaviour at the origin.
This suggests that the process of excitation
 for a charged exciton at fixed angular
 momentum $M$ involves repetition of the 
two-step process corresponding to
 Figs.7(b) and 7(d). Each cycle pushes
 the peaks to higher r and forces the
 RPDF to vanish more quickly at the origin.

\section{Summary}

To summarize, we have studied a negatively
 charged exciton in a uniform
 B-field by means of numerical diagonalization.
 We have discussed our results
 in terms of the PL spectra for various
 in-plane confinements and
 experimentally relevant B-fields. The
 PL spectra were calculated
 considering recombination from a range
 of initial angular momentum
 states of the charged exciton ($-3\le 
M \le0$). A realistic form factor
  consistent with 200--300\AA \
 quantum wells
 has been used to modify the interactions.
 We found that our weak in-plane confinement
 results agree
 very closely with published experimental
 spectra for high
 mobility quantum wells 
 (Ref.~\cite{Shields2}). We also get
 good agreement with the PL
 energies and intensities of published
 shakeup lines in the PL spectra
 (Ref.~\cite{FinkelsteinSU}). These results
 demonstrate that the 
expected PL signatures of charged
 excitons are not qualitatively
 affected by the
 presence of a range of initial angular
 momentum states. The one exception
 to this is the triplet line which appears 
at finite B-field. This is 
due to the $M=0$ triplet charged exciton
 which must be present for the
 line to be observed. We also discussed
 the PL spectra (including shakeup lines)
 for the cases of intermediate and strong
 in-plane confinements.
 These represent predictions of the 
expected PL spectra from 
charged excitons in quantum dots.
  Our calculated spectra show 
interesting structure which should be
 observable and we hope this
 will stimulate further experimental
 study on such systems.
 Finally, we calculated various 
electron-hole radial pair distribution
 functions for the charged excitons.
 These allowed us to examine in detail 
the structure of the charged exciton
 states as a function of angular momentum,
 B-field and in-plane
 confinement. The excited states of the 
charged exciton were also studied.
 We found that excitation occurs via two
 methods. First the charged
 exciton can undergo partial ionization
 whereby one of the electrons
 moves away from the hole leaving a
 configuration of an exciton
 orbited by an outer electron. Second,
 both electrons remain equally
 close to the hole, but their 
correlations with it change. The
 combination of these two processes
 provides insight into the excitation
 mechanism for charged excitons at
 a given angular momentum.

We would like to thank A.J. Shields 
and G.Finkelstein for
 preprints. Funding was provided by EPSRC 
through direct support
 (J.R.C.) and through Research Grant
 No. GR/K 15619 (N.F.J. and V.N.N.).

\newpage \centerline{\bf Figure Captions}

\bigskip

\noindent Figure 1: PL Spectra for a
 weak in-plane confinement and rod length
 $L=3l$ (240\AA \ at B=10T).
 Recombination from the lowest energy
 singlet ($X^-_s$) and
 triplet ($X^-_t$) charged excitons with 
 $M=-3,-2,-1,0$  is included.
 Recombination from the exciton ($X$) ground
 state ($M=0$) is also shown. A Gaussian broadening 
of 0.3meV FWHM is included. 
\bigskip

\noindent Figure 2: PL Spectra for an
 intermediate in-plane confinement and
 rod length $L=3l$. Recombination from 
the lowest energy singlet ($X^-_s$) and
 triplet ($X^-_t$) charged excitons with $M=-3,-2,-1,0$
 is included. Recombination from the
 exciton ($X$) ground state ($M=0$) is
 shown for reference. 
A Gaussian broadening
 of 0.3meV FWHM is included.
\bigskip

\noindent Figure 3: PL Spectra for a
 strong in-plane confinement and rod 
length $L=3l$. Recombination from the
 lowest energy singlet ($X^-_s$) and
 triplet ($X^-_t$) charged excitons with $M=-3,-2,-1,0$
 is included. Recombination from the
 exciton ($X$) ground state ($M=0$) 
is shown for reference. 
A Gaussian broadening
 of 0.3meV FWHM is included.
\bigskip

\noindent Figure 4: Shakeup lines in 
the PL spectrum for a weak in-plane 
confinement and rod length $L=3l$. (a)
 ${\rm SU_1(n=1)}$ shakeup from the
 lowest energy $M=-3,-2,-1,0$ singlet
 and triplet charged exciton states. 
(b) ${\rm SU_1(n=0)}$ shakeup from lowest
 energy $M=1$ singlet and triplet
 charged exciton states. The non-shakeup
 spectra of Fig.1 are also shown for
 reference. A Gaussian broadening of
 0.3meV FWHM is included.
\bigskip

\noindent Figure 5: Shakeup lines in
 the PL spectrum for an intermediate
 in-plane confinement and rod length $L=3l$.
 (a) ${\rm SU_1(n=1)}$ shakeup
 from the lowest energy $M=-3,-2,-1,0$
 singlet and triplet charged exciton
 states. (b) ${\rm SU_1(n=0)}$ shakeup
 from lowest energy $M=1$ singlet and
 triplet charged exciton states; note
 that different magnifications are 
used for the singlet and triplet processes.
 The non-shakeup spectra of Fig.2
 are also shown for reference. 
A Gaussian broadening of 0.3meV FWHM is included.
\bigskip

\noindent Figure 6: Shakeup lines in the
 PL spectrum for a strong in-plane
 confinement and rod length $L=3l$. (a)
 ${\rm SU_1(n=1)}$ shakeup from the 
lowest energy $M=-3,-2,-1,0$ singlet and
 triplet charged exciton states. 
(b) ${\rm SU_1(n=0)}$ shakeup from lowest
 energy $M=1$ singlet and triplet
 charged exciton states; note that different
 magnifications are used for
 the singlet and triplet processes. The
 non-shakeup spectra of Fig.3 are also
 shown for reference. A Gaussian broadening
 of 0.3meV FWHM is included.
\bigskip

\noindent Figure 7: Schematic diagram
 showing radial pair distribution 
functions (RPDF) for typical configurations
 (a) Both electrons equally correlated
 with the hole.
 (b) The electron--hole radial pair
 distriution function $\rho(r)$
 associated with
 part (a). (c) One electron closer to the
 hole than the other. (d) The
 electron--hole radial pair 
distribution function $\rho(r)$ 
associated with part (c).
\bigskip

\noindent Figure 8: Electron-hole radial
 pair distribution function $\rho(r)$
 at various B-field for lowest energy $M=0$ 
singlet charged exciton in a weak in-plane
 confinement and rod length $L=3l$.
\bigskip

\noindent Figure 9: Electron-hole radial 
pair distribution function $\rho(r)$
 for lowest energy  
charged exciton states with B=10T in a 
weak in--plane confinement and
 rod length $L=3l$. Charged exciton states with
 $M=-3 \rightarrow 3$ are considered.
 a) Singlet states. b) Triplet states.
\bigskip

\noindent Figure 10: Electron-hole radial
 pair distribution function $\rho(r)$
 for lowest energy 
 charged exciton states with B=10T in
 a strong in--plane confinement
 and rod length $L=3l$. Charged exciton states with
 $M=-3 \rightarrow 3$ are considered.
 a) Singlet states. b) Triplet states.
\bigskip

\noindent Figure 11: Electron-hole radial
 pair distribution function $\rho(r)$
 for the lowest five energy states of (a)
 $M=0$ singlet charged exciton (level 1 
is the singlet ground state) and (b)
 $M=-1$ triplet charged exciton (level 1
 is the triplet ground state).
 B=10T with a weak in-plane confinement
 and a rod length $L=3l$.
\bigskip

\noindent Figure 12: Electron-hole radial
 pair distribution function $\rho(r)$
 for the lowest five energy states of an
 $M=-2$ singlet charged exciton in a strong
 in-plane confinement with B=10T
 and a rod length $L=3l$.
\bigskip

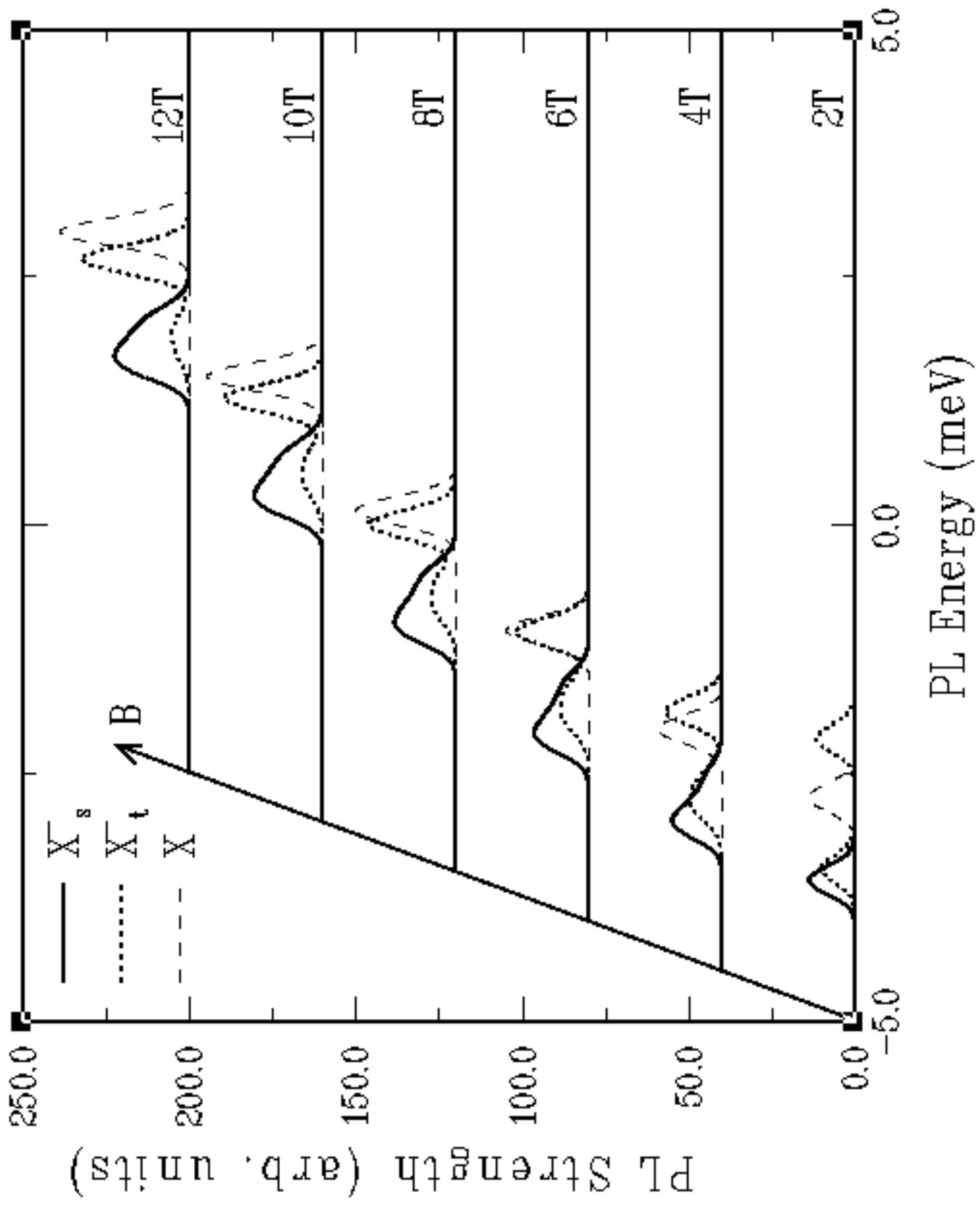
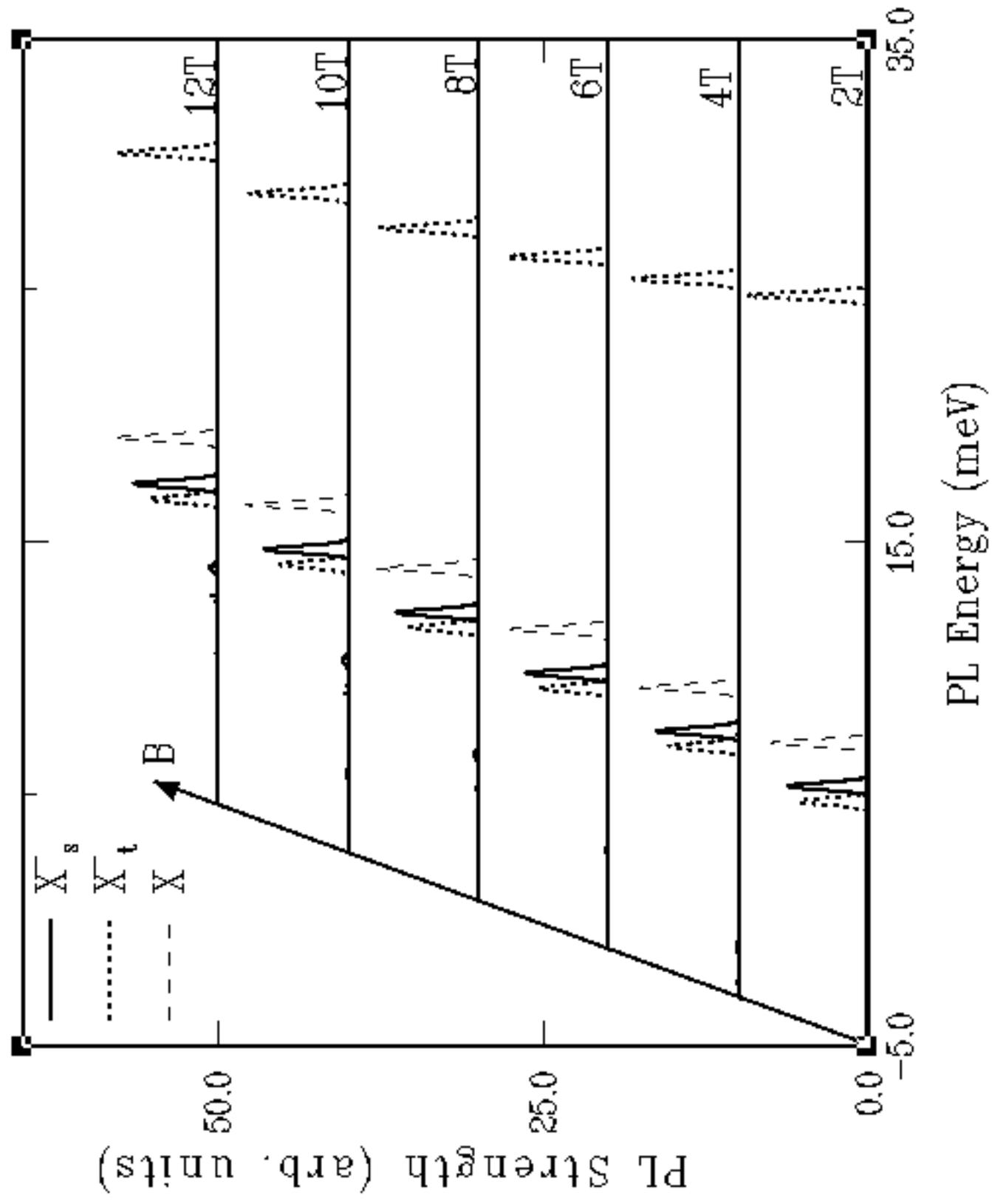
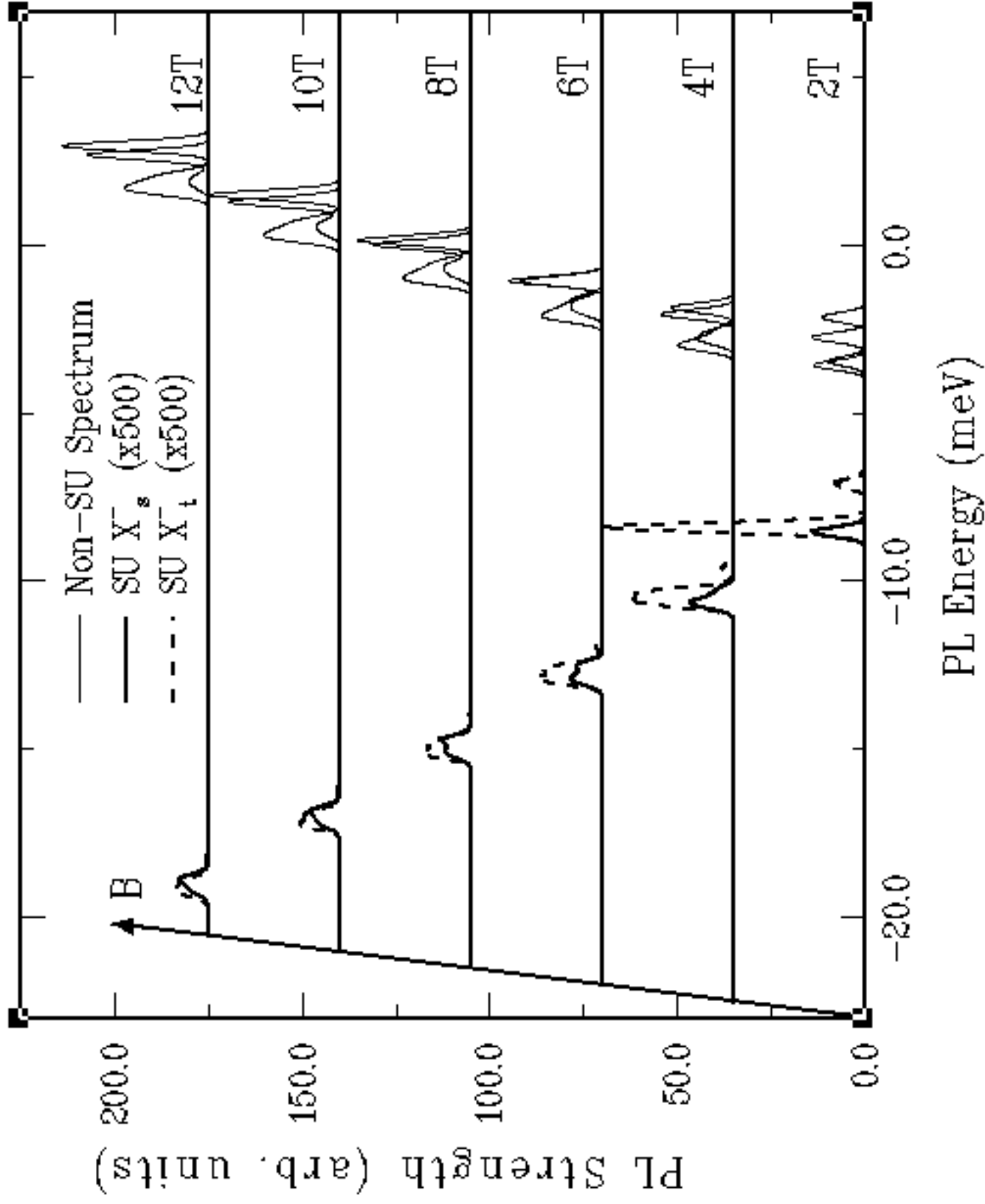
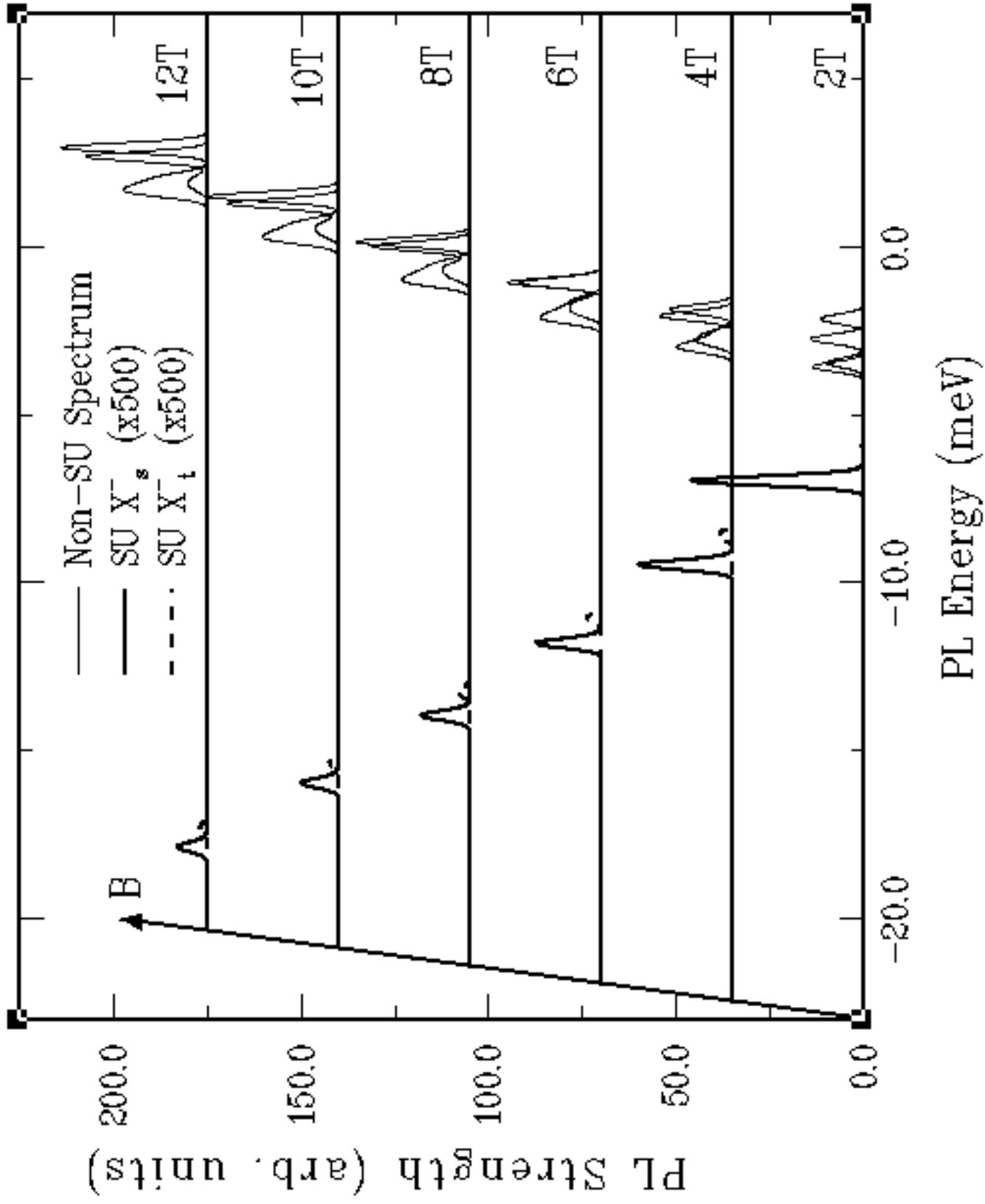
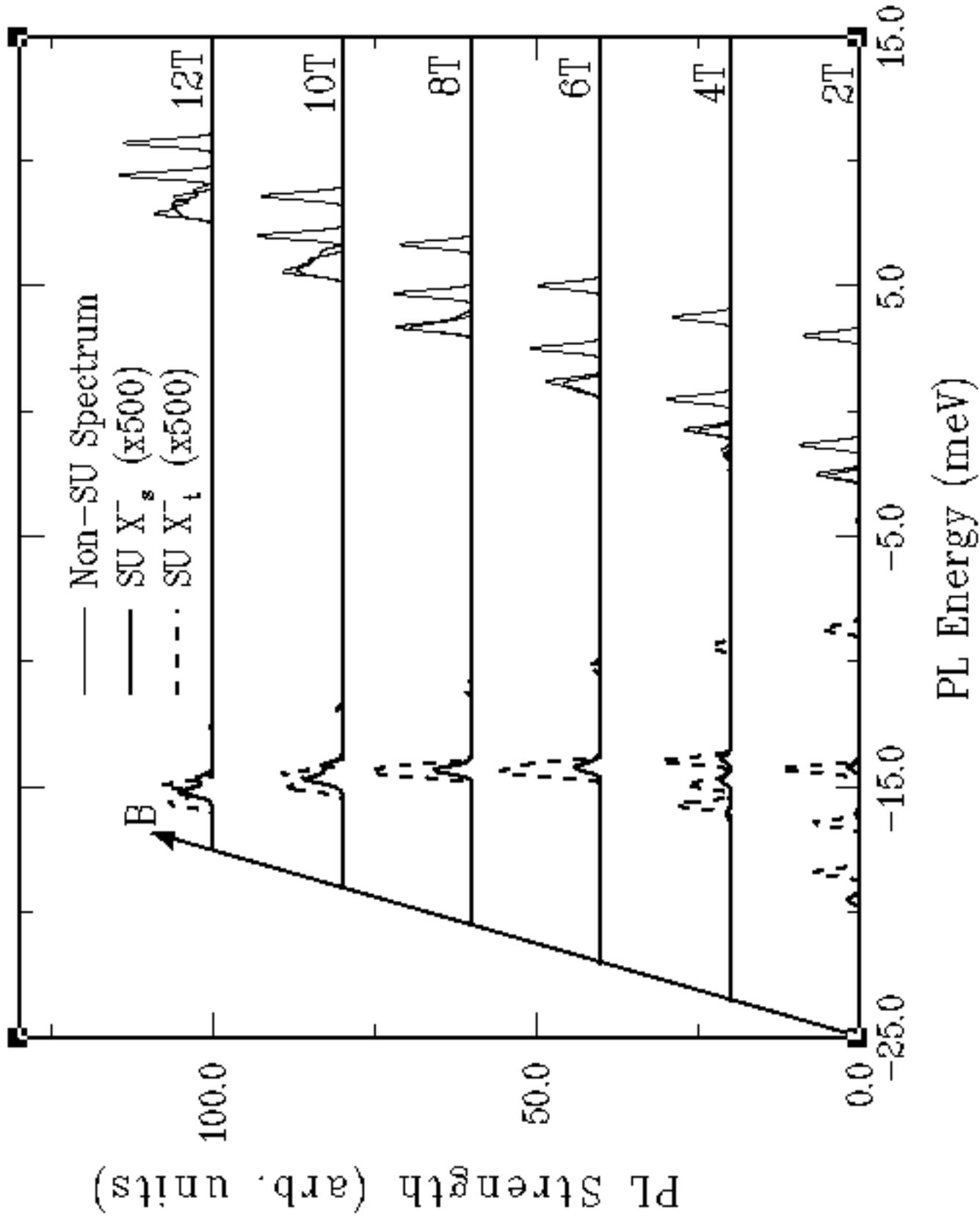
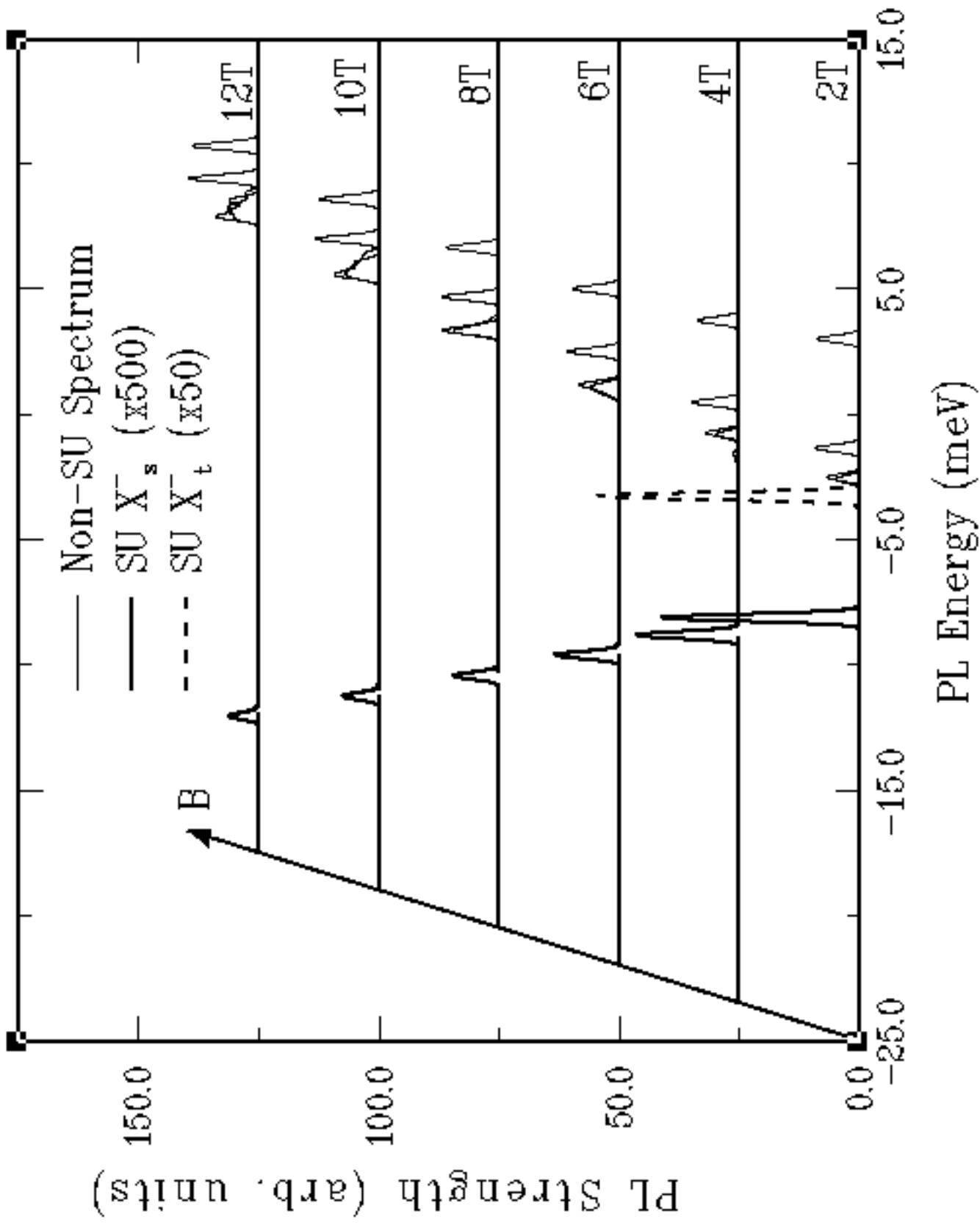
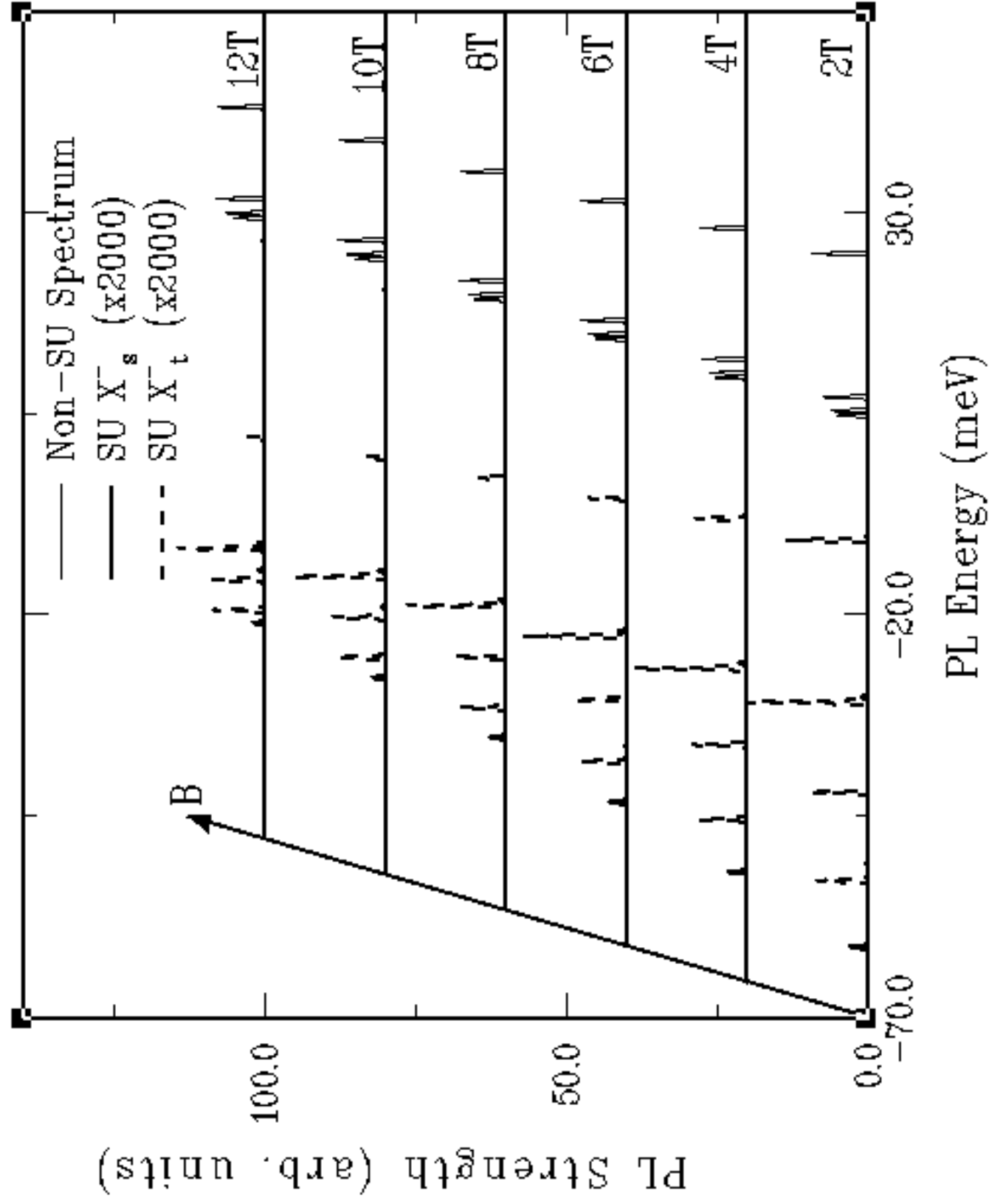
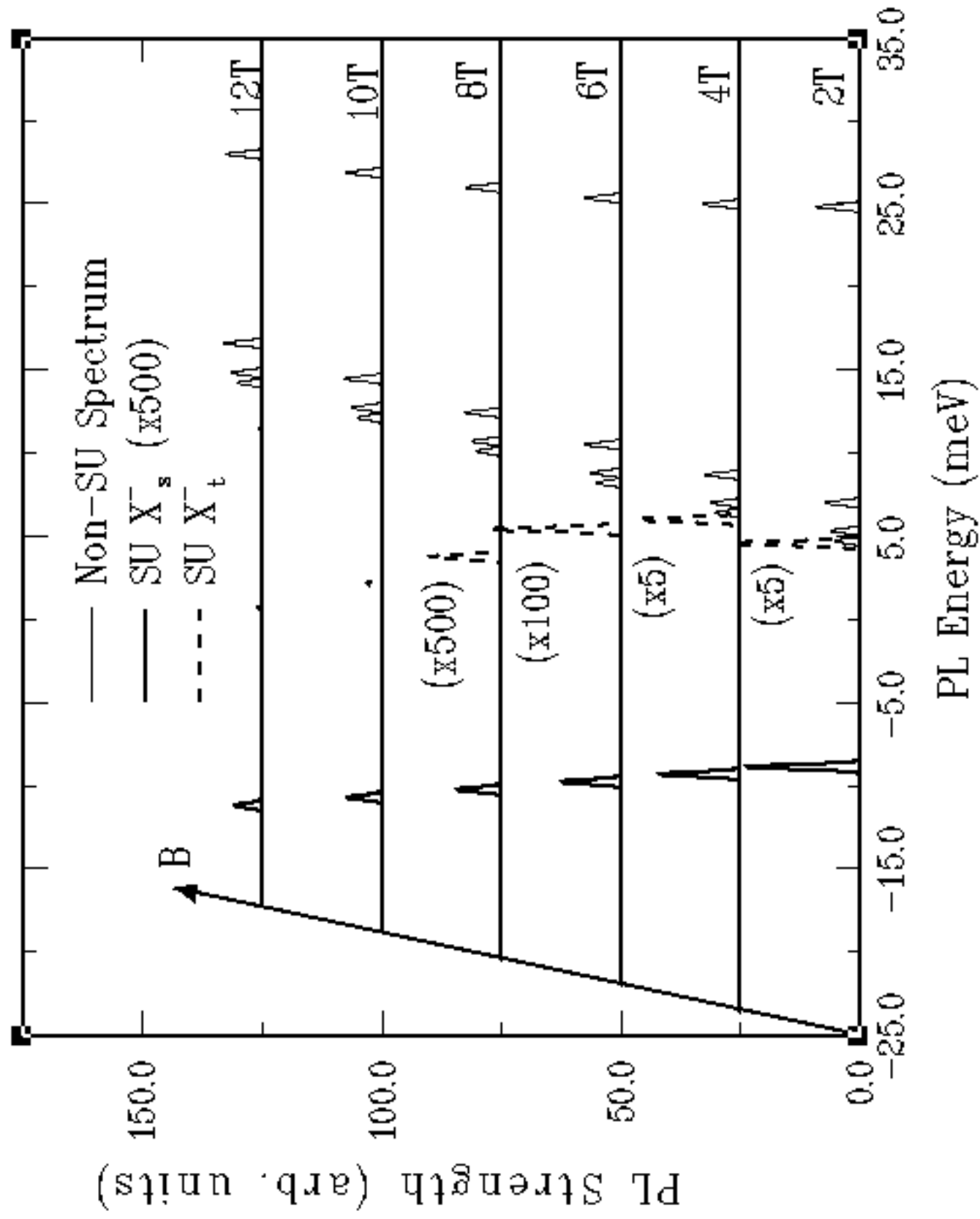
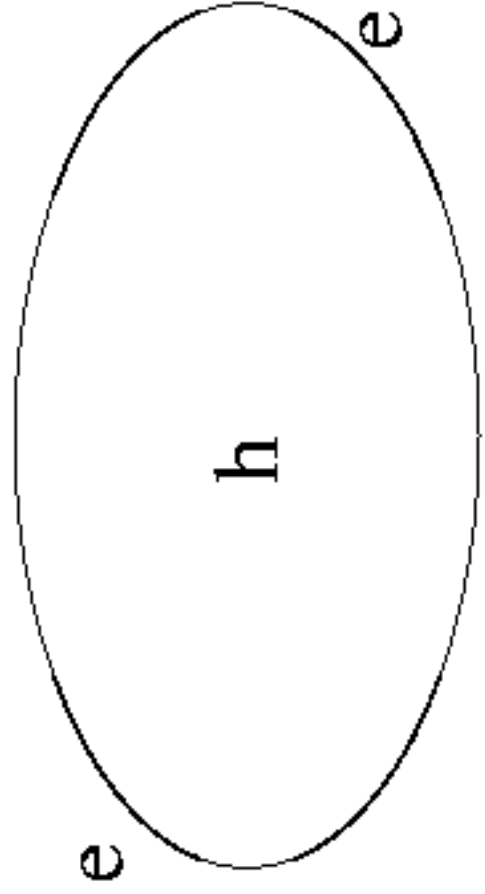
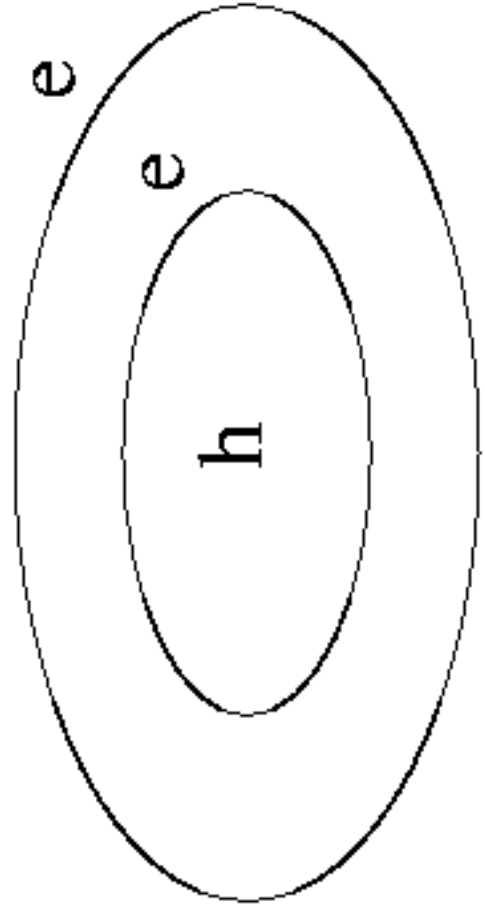
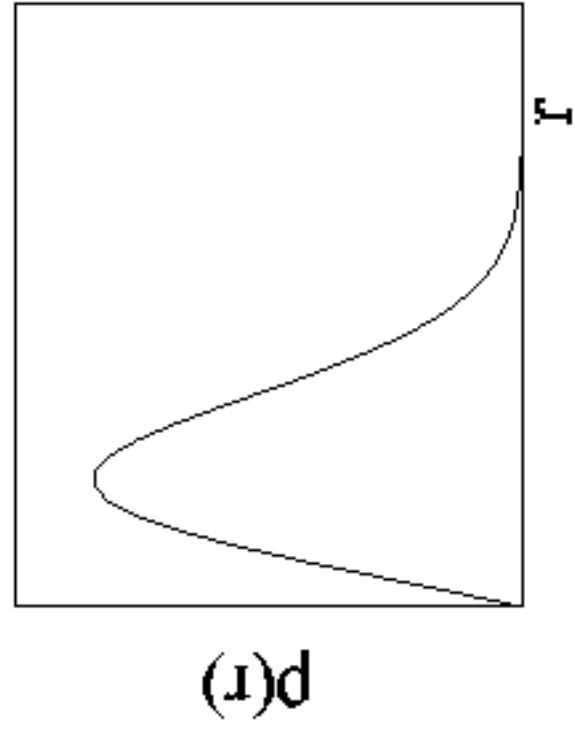
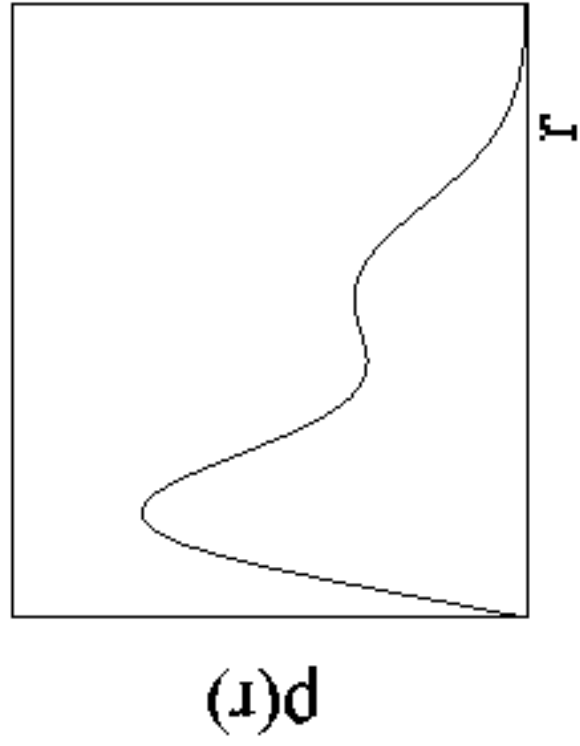
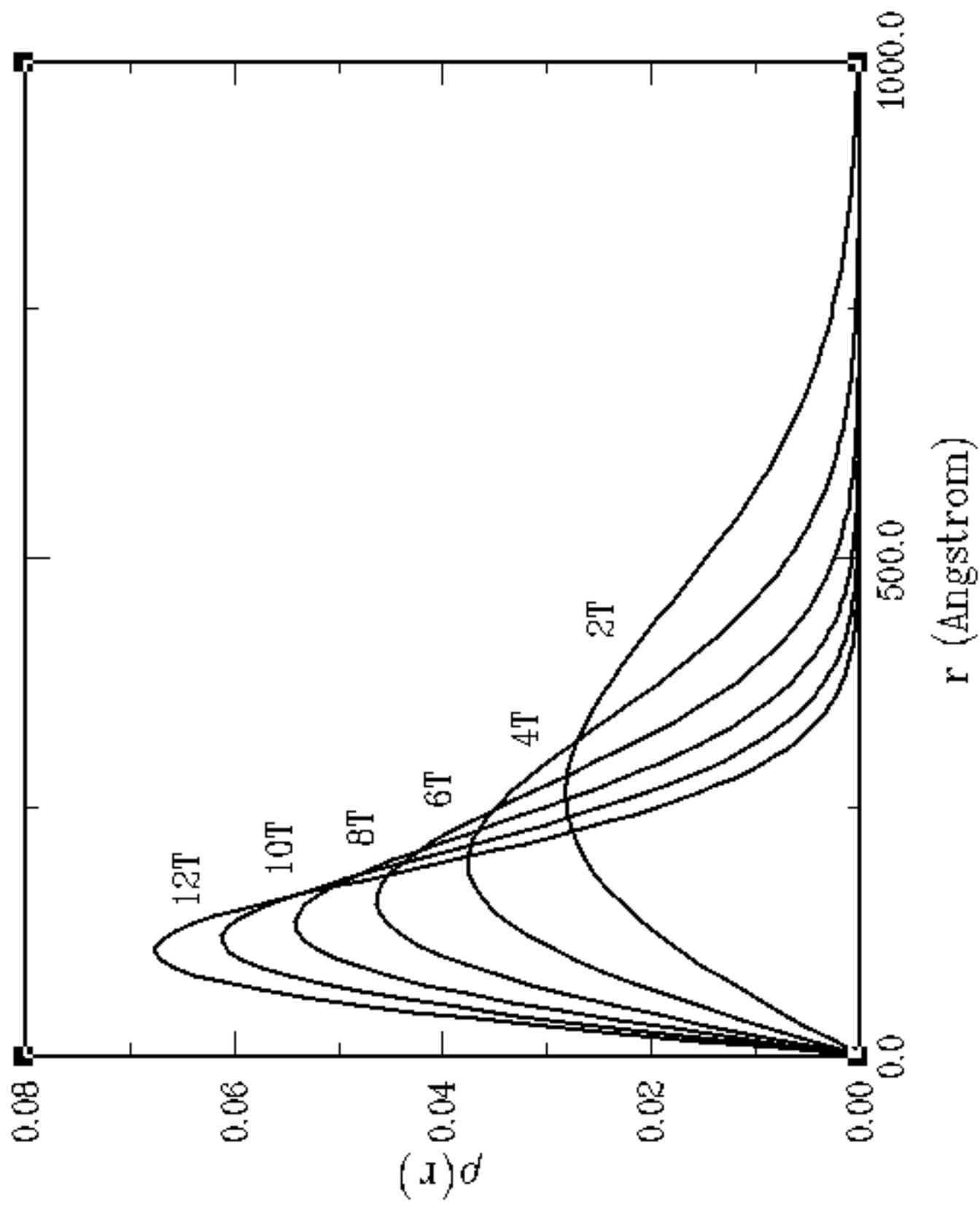
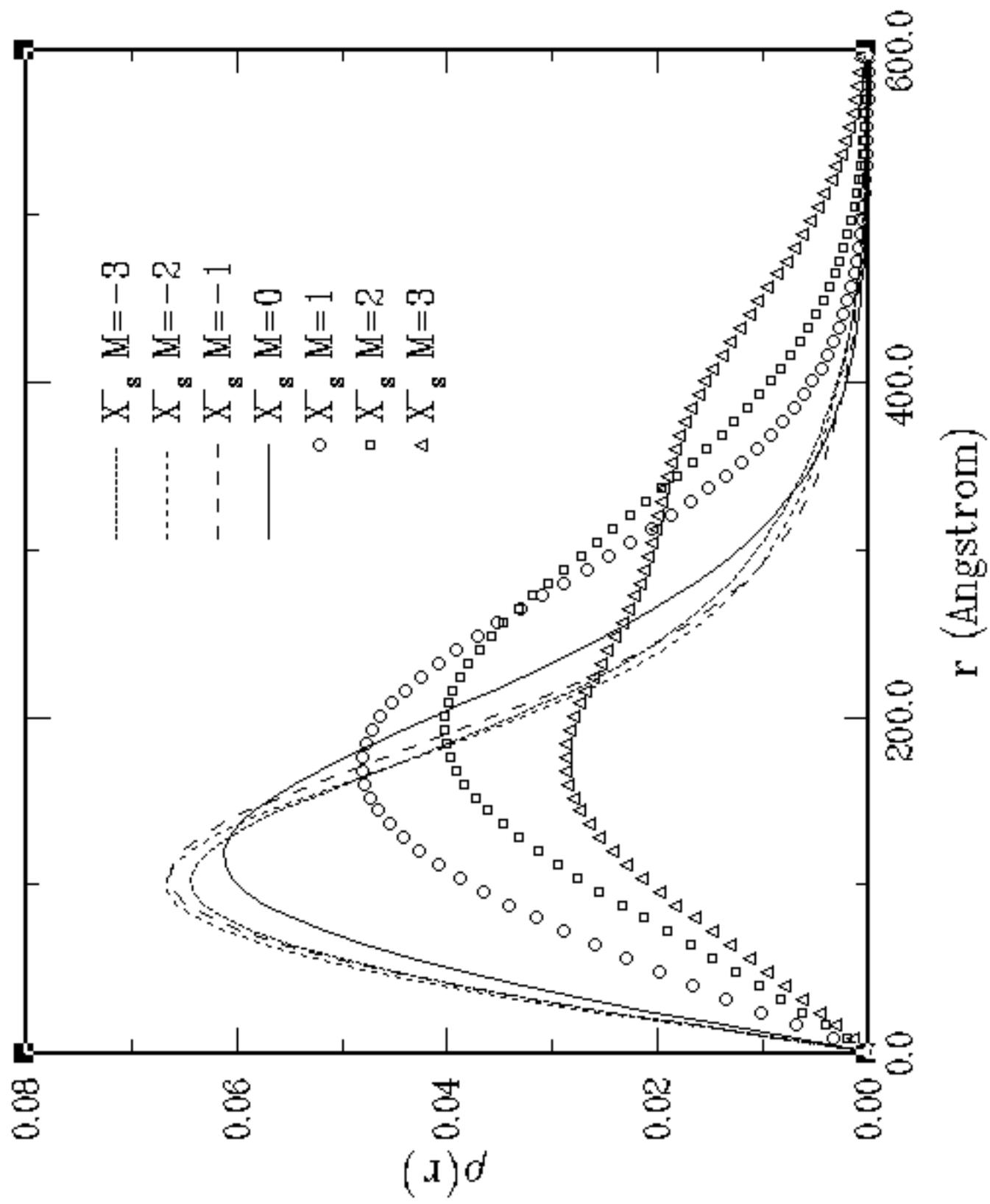
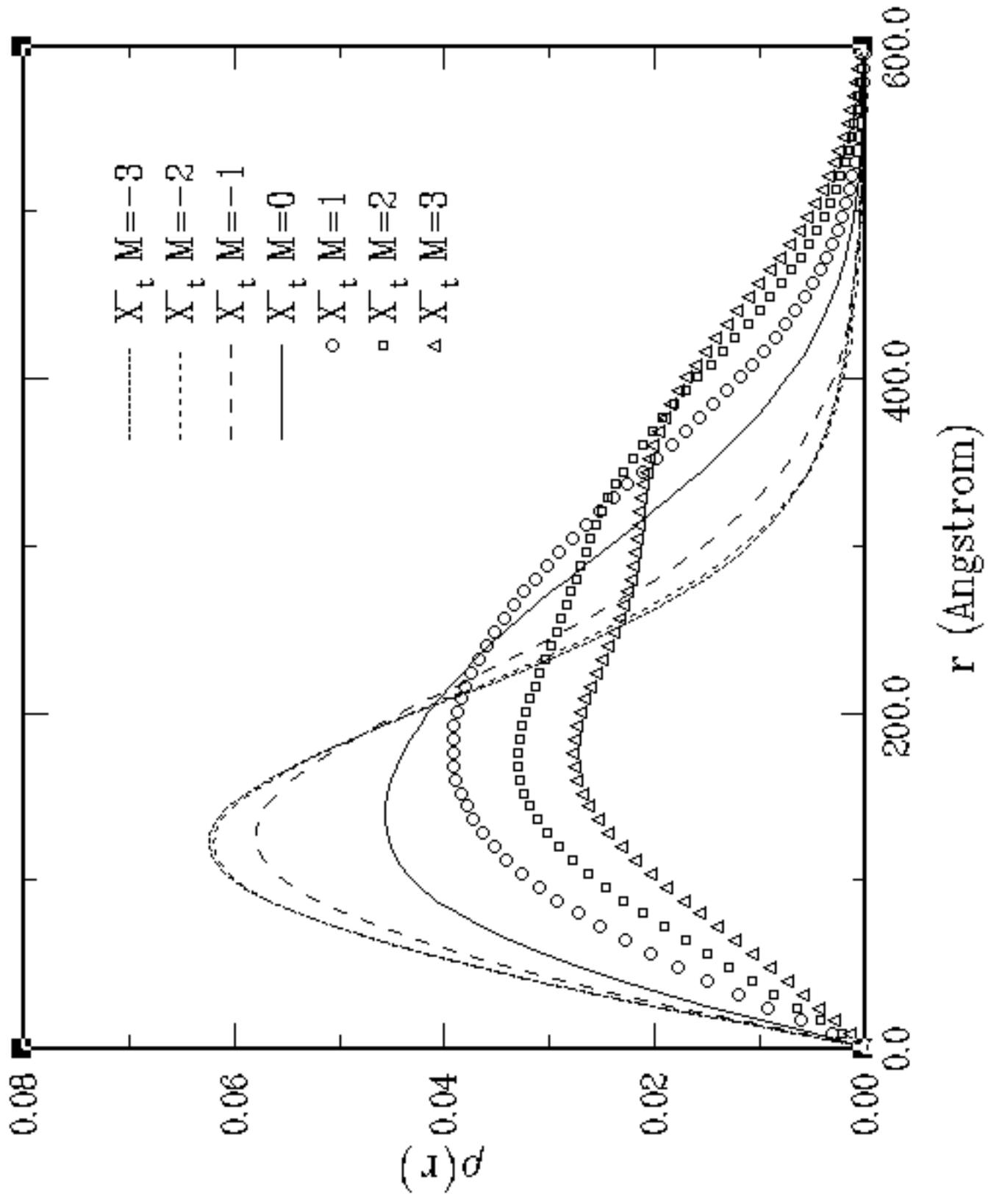
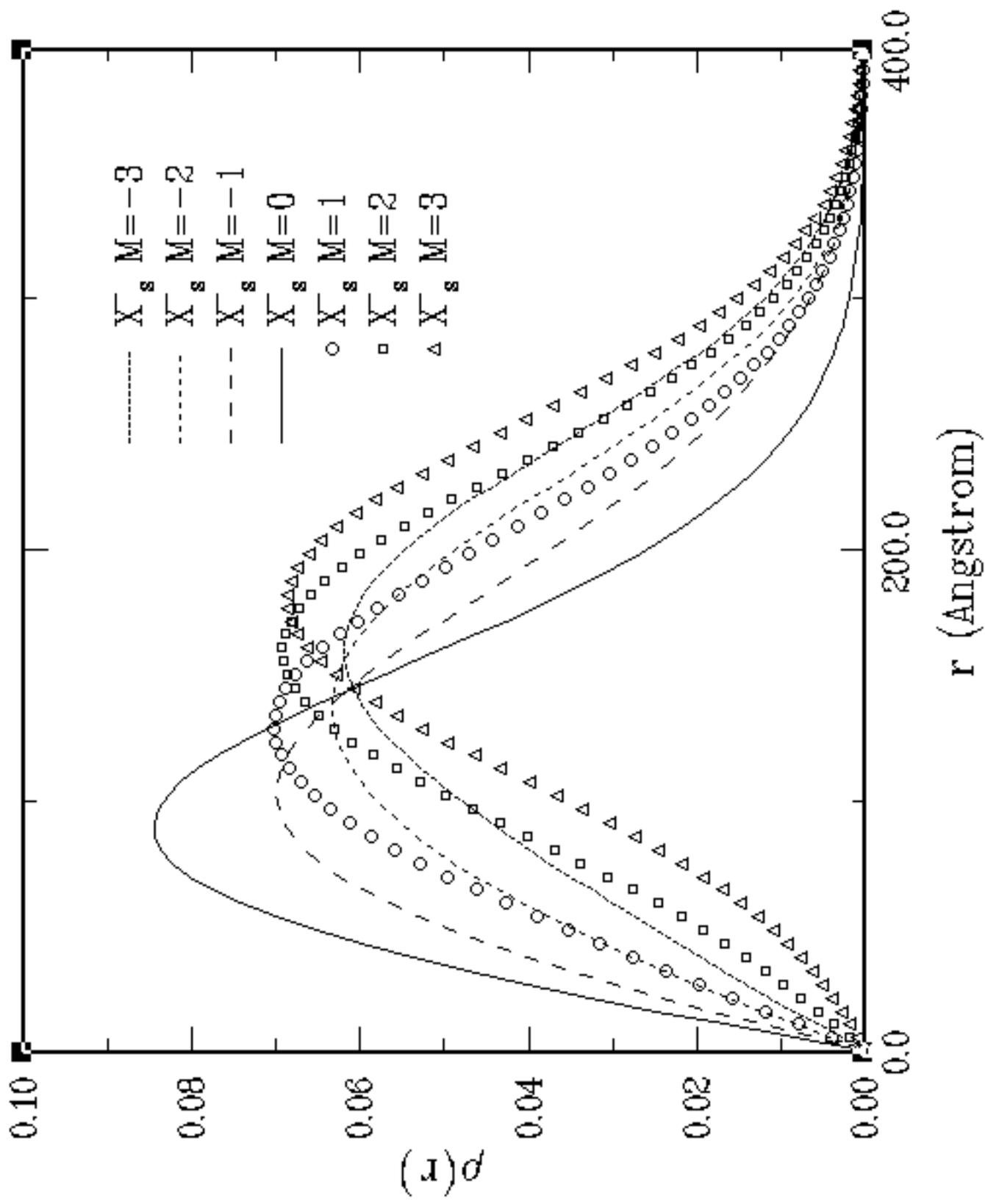
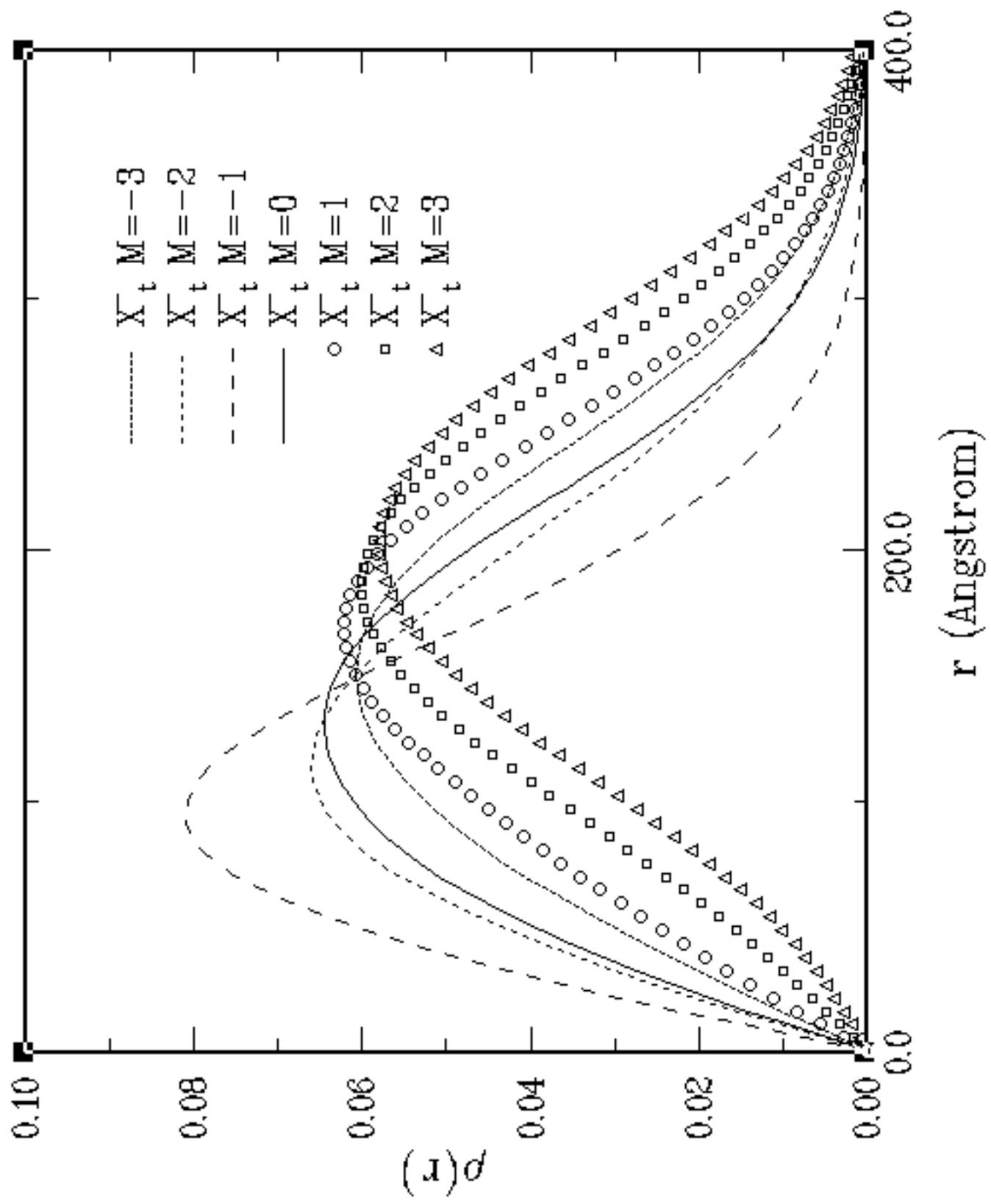
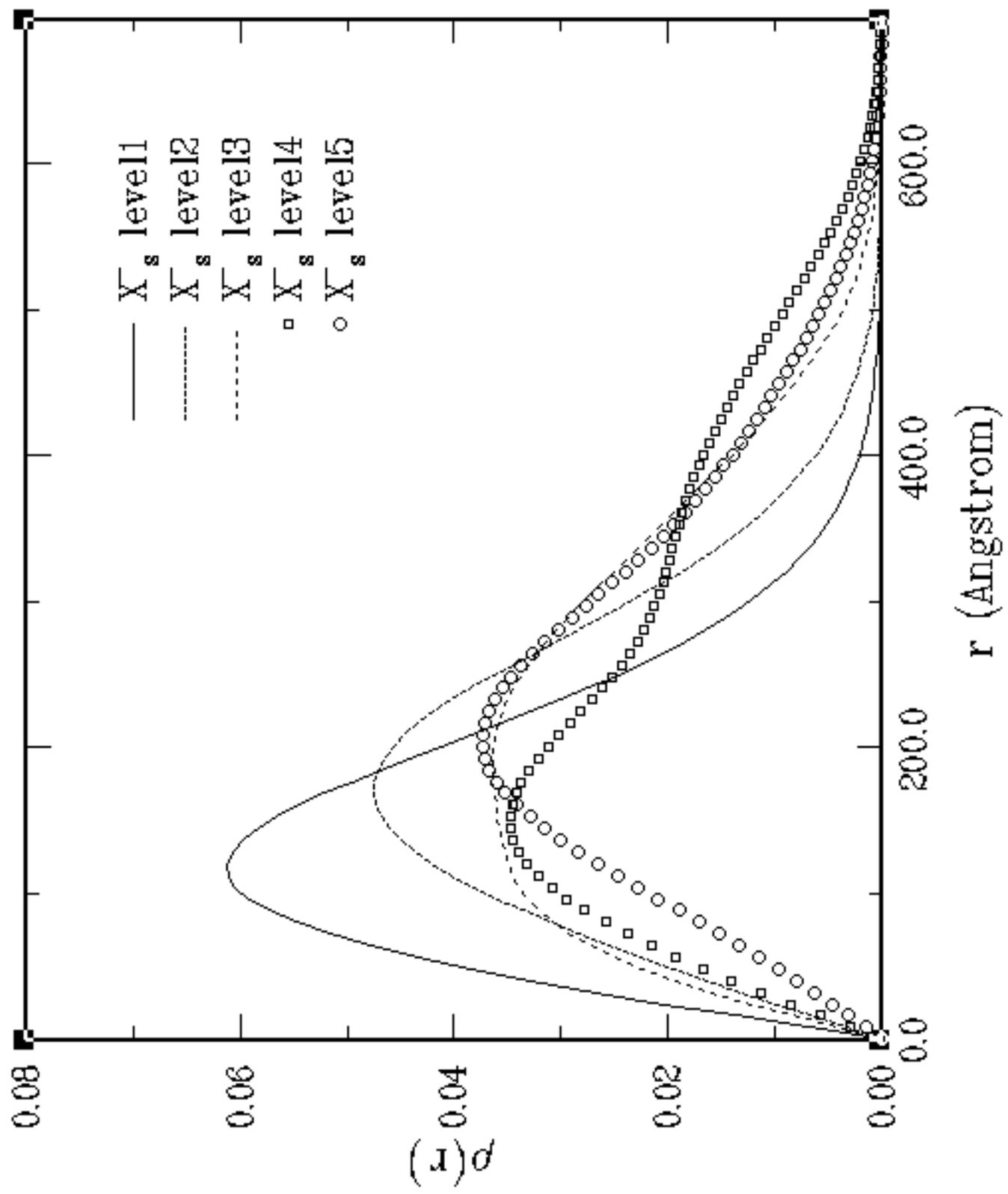
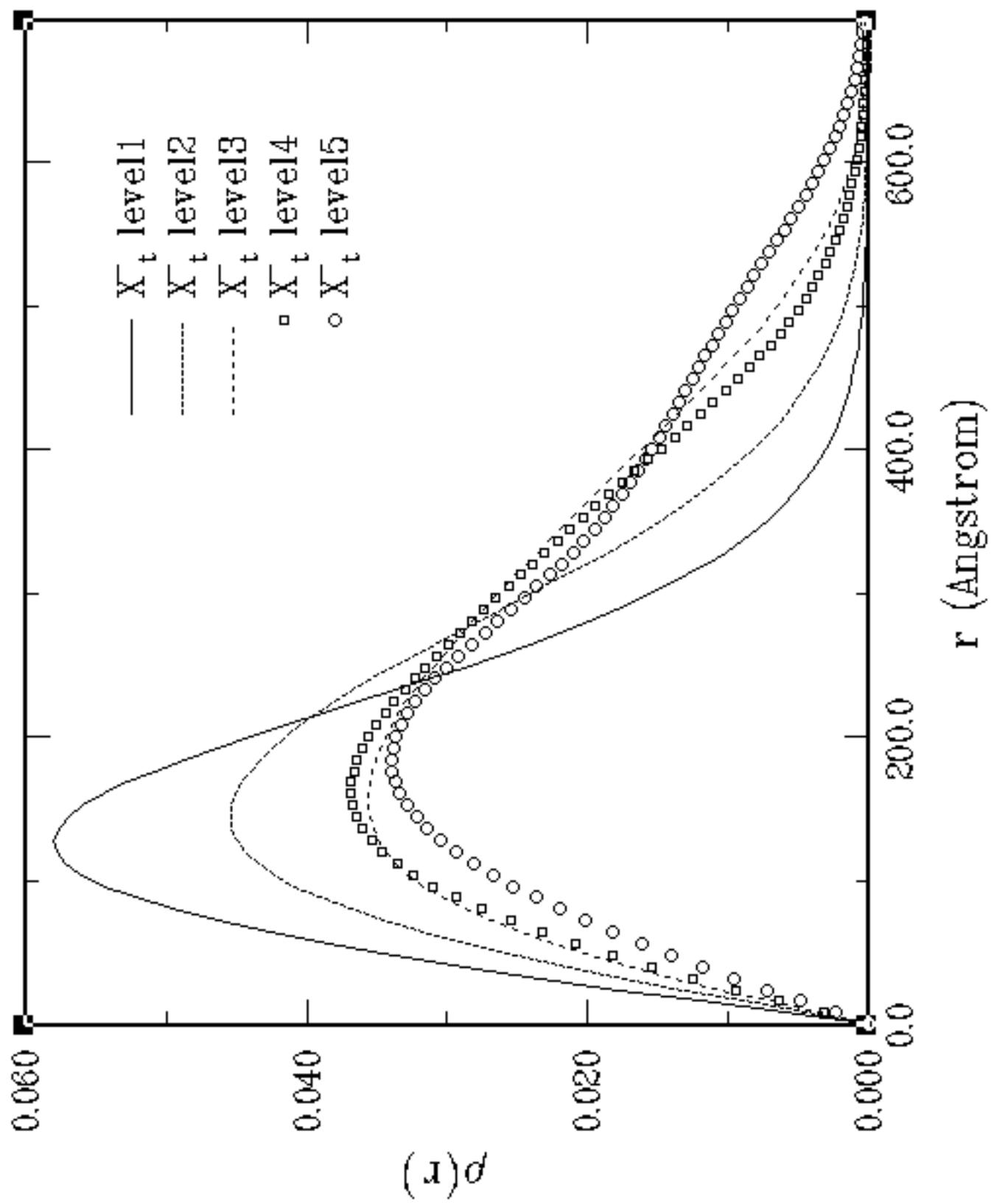
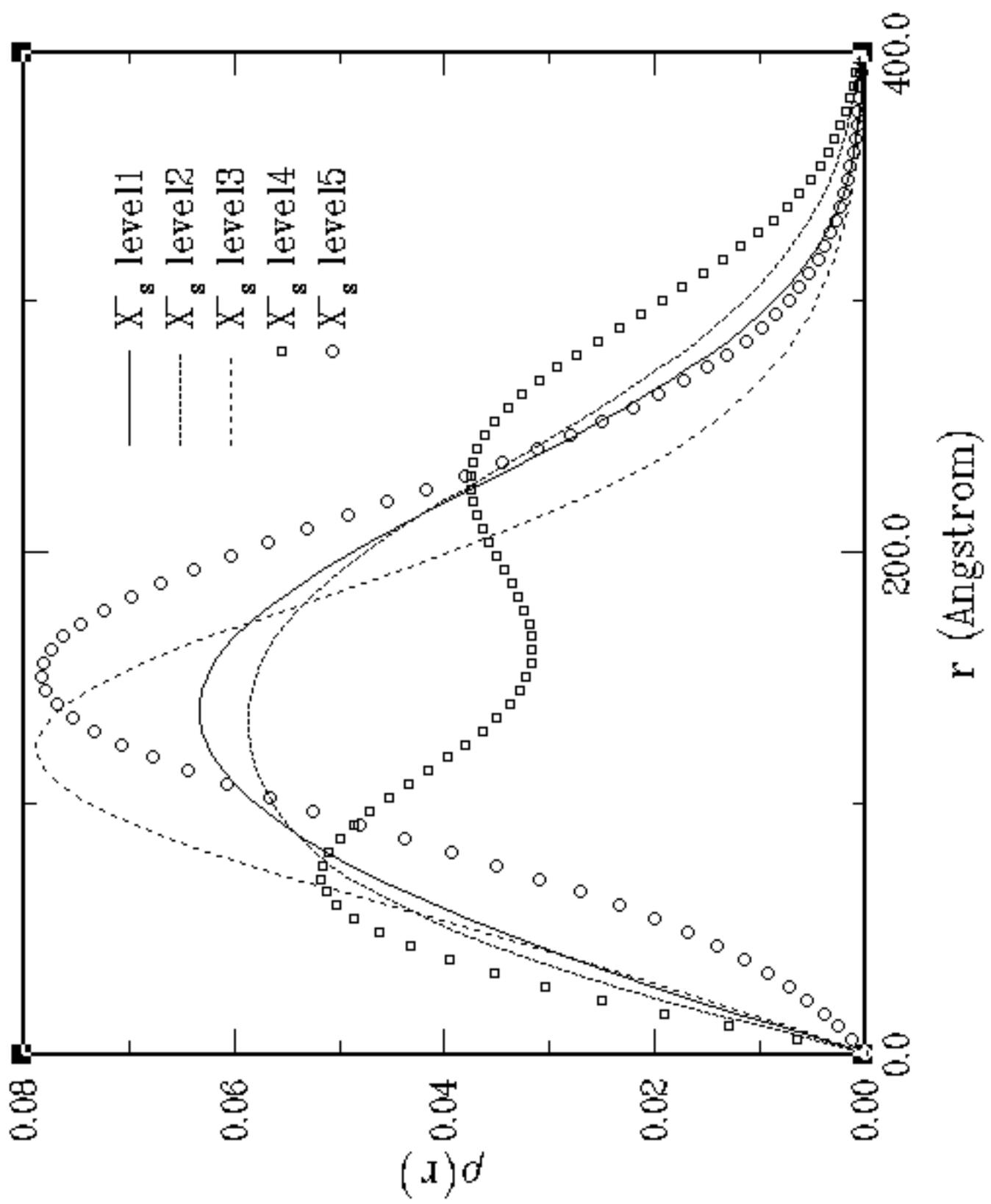
\end{document}